\documentclass[fleqn,usenatbib,useAMS,twocolumn]{aastex62}
\usepackage[utf8]{inputenc}
\usepackage{todonotes}
\usepackage{color}
\usepackage{graphicx}
\usepackage{amsmath}
\usepackage{amssymb}
\usepackage{bm}
\usepackage{acronym}
\usepackage{ifthen}
\usepackage[normalem]{ulem}
\usepackage{hyperref}
\usepackage{etoolbox}
\usepackage{fancyhdr}
\usepackage{xspace}
\usepackage{textcomp}
\usepackage{multirow}
\usepackage[caption=false]{subfig}
\usepackage{lineno}
\usepackage{tabularx}
\usepackage{longtable}
\usepackage{afterpage}
\usepackage{xspace}
\usepackage{graphicx, verbatim, ifthen}

\reportnum{LIGO-P2100185}

\newcommand\hu{~\rm{km \, s^{-1}} \,Mpc^{-1}\xspace}

\newcommand{\gammamax}{12.0\xspace}
\newcommand{\gammamin}{0.0\xspace}
\newcommand{\kappamax}{6.0\xspace}
\newcommand{\kappamin}{0.0\xspace}
\newcommand{\zpmax}{4.0\xspace}
\newcommand{\zpmin}{0.0\xspace}
\newcommand{\Hnotmax}{200.0\xspace}
\newcommand{\Hnotmin}{10.0\xspace}
\newcommand{\Omnotmax}{1.0\xspace}
\newcommand{\Omnotmin}{0.0\xspace}
\newcommand{\wnotmax}{0.0\xspace}
\newcommand{\wnotmin}{-3.0\xspace}
\newcommand{\Rnotmax}{100.0\xspace}
\newcommand{\Rnotmin}{0.0\xspace}
\newcommand{\BPLalphaonemax}{12.0\xspace}
\newcommand{\BPLalphaonemin}{1.5\xspace}
\newcommand{\BPLalphatwomax}{12.0\xspace}
\newcommand{\BPLalphatwomin}{1.5\xspace}
\newcommand{\BPLbetamax}{12.0\xspace}
\newcommand{\BPLbetamin}{-4.0\xspace}
\newcommand{\BPLmmaxmax}{200.0\xspace}
\newcommand{\BPLmmaxmin}{50.0\xspace}

\newcommand{\BPLmminmin}{2.0\xspace}
\newcommand{\BPLbmax}{1.0\xspace}
\newcommand{\BPLbmin}{0.0\xspace}
\newcommand{\BPLdeltammax}{10.0\xspace}
\newcommand{\BPLdeltammin}{0.0\xspace}
\newcommand{\PLGalphamax}{12.0\xspace}
\newcommand{\PLGalphamin}{1.5\xspace}
\newcommand{\PLGbetamax}{12.0\xspace}
\newcommand{\PLGbetamin}{-4.0\xspace}
\newcommand{\PLGmmaxmax}{200.0\xspace}
\newcommand{\PLGmmaxmin}{50.0\xspace}
\newcommand{\PLGmminmax}{10.0\xspace}
\newcommand{\PLGmminmin}{2.0\xspace}
\newcommand{\PLGmugmax}{50.0\xspace}
\newcommand{\PLGmugmin}{20.0\xspace}
\newcommand{\PLGsigmagmax}{10.0\xspace}
\newcommand{\PLGsigmagmin}{0.4\xspace}
\newcommand{\PLGlambdapeakmax}{1.0\xspace}
\newcommand{\PLGlambdapeakmin}{0.0\xspace}
\newcommand{\PLGdeltammax}{10.0\xspace}
\newcommand{\PLGdeltammin}{0.0\xspace}
\newcommand{\PLalphamax}{12.0\xspace}
\newcommand{\PLalphamin}{1.5\xspace}
\newcommand{\PLbetamax}{12.0\xspace}
\newcommand{\PLbetamin}{-4.0\xspace}
\newcommand{\PLmmaxmax}{200.0\xspace}
\newcommand{\PLmmaxmin}{50.0\xspace}
\newcommand{\PLmminmax}{10.0\xspace}
\newcommand{\PLmminmin}{2.0\xspace}
\newcommand{\zbest}{0.01\xspace}
\newcommand{\Hzbest}{62^{+87}_{-41\xspace}\xspace}

\newcommand{\HzbestBNS}{74^{+34}_{-13\xspace}\xspace}
\newcommand{\NevicarogwSNRten}{60\xspace}
\newcommand{\NevicarogwSNRtwelve}{35\xspace}
\newcommand{\Nevicarogw}{42\xspace}
\newcommand{\Nevgwcosmo}{47\xspace}
\newcommand{\PLGBPLlogtenfact}{-0.5\xspace}
\newcommand{\PLGPLGlogtenfact}{0.0\xspace}
\newcommand{\PLGPLlogtenfact}{-1.9\xspace}

\newcommand{\BPLBPLlogtenBfactCOSMO}{-0.4\xspace}
\newcommand{\PLGPLGlogtenBfactCOSMO}{-0.3\xspace}
\newcommand{\PLPLlogtenBfactCOSMO}{0.2\xspace}
\newcommand{\HnoticarogwBPL}{44^{+52}_{-24}\xspace}

\newcommand{\HnotnoticarogwwithBNSlogBPL}{68^{+13}_{-8}\xspace}

\newcommand{\HnoticarogwimproGWTConeBPL}{12\xspace}
\newcommand{\HnoticarogwPLG}{50^{+37}_{-30}\xspace}

\newcommand{\HnotnoticarogwwithBNSlogPLG}{68^{+12}_{-8}\xspace}

\newcommand{\HnoticarogwimproGWTConePLG}{17}
\newcommand{\HnoticarogwPL}{109^{+43}_{-54}\xspace}

\newcommand{\HnotnoticarogwwithBNSlogPL}{69^{+21}_{-8}\xspace}

\newcommand{\mugcarogw}{32^{+6}_{-8}\xspace}
\newcommand{\HnoticarogwSYMBPL}{66^{+98}_{-47}\xspace}

\newcommand{\HnotnoticarogwwithBNSlogSYMBPL}{73^{+34}_{-14}\xspace}
\newcommand{\HnoticarogwSYMPLG}{62^{+90}_{-42}\xspace}

\newcommand{\HnotnoticarogwwithBNSlogSYMPLG}{72^{+30}_{-13}\xspace}
\newcommand{\HnoticarogwSYMPL}{104^{+74}_{-77}\xspace}

\newcommand{\HnotnoticarogwwithBNSlogSYMPL}{79^{+44}_{-19}\xspace}

\newcommand{\HnotgwcosmoGWnof}{67^{+46}_{-28}\xspace}

\newcommand{\HnotgwcosmobJband}{67^{+14}_{-12}\xspace}

\newcommand{\HnotgwcosmoKband}{67^{+13}_{-12}\xspace}

\newcommand{\HnotgwcosmoEmpty}{67^{+14}_{-13}\xspace}

\newcommand{\HnotgwcosmoBNSbJband}{68^{+9}_{-6}\xspace}

\newcommand{\HnotgwcosmoBNSKband}{68^{+8}_{-6}\xspace}

\newcommand{\HnotgwcosmoSYMbJband}{69^{+25}_{-23}\xspace}

\newcommand{\HnotgwcosmoSYMKband}{68^{+25}_{-21}\xspace}

\newcommand{\HnotgwcosmoBNSSYMbJband}{71^{+18}_{-12}\xspace}

\newcommand{\HnotgwcosmoBNSSYMKband}{70^{+17}_{-13}\xspace}

\newcommand{\HnotgwcosmoimproGWTConeKband}{42}

\newcommand{\HnotgwcosmoimproBNSKband}{44}

\newcommand{\PL}{\textsc{Truncated}\xspace}
\newcommand{\PLG}{\textsc{Power Law + Peak}\xspace}
\newcommand{\BPL}{\textsc{Broken Power Law}\xspace}

\newcommand{\gladeplus}{\texttt{GLADE+}\xspace}
\newcommand{\glade}{\texttt{GLADE}\xspace}

\newcommand{\mmin}{\xspace m_{\rm min}\xspace}
\newcommand{\mmax}{\xspace m_{\rm max}\xspace}
\newcommand{\Msol}{\xspace\rm{M}_{\odot}\xspace}

\shorttitle{Constraints on the cosmic expansion history from GWTC--3}

\shortauthors{Abbott et al.}

\begin{document}
\title{Constraints on the cosmic expansion history from  GWTC--3}

\newtoggle{fullauthorlist}
\togglefalse{fullauthorlist}
\toggletrue{fullauthorlist}
\iftoggle{fullauthorlist}{
\author{R.~Abbott}
\affiliation{LIGO Laboratory, California Institute of Technology, Pasadena, CA 91125, USA}
\author{H.~Abe}
\affiliation{Graduate School of Science, Tokyo Institute of Technology, Meguro-ku, Tokyo 152-8551, Japan  }
\author{F.~Acernese}
\affiliation{Dipartimento di Farmacia, Universit\`a di Salerno, I-84084 Fisciano, Salerno, Italy  }
\affiliation{INFN, Sezione di Napoli, Complesso Universitario di Monte S. Angelo, I-80126 Napoli, Italy  }
\author[0000-0002-8648-0767]{K.~Ackley}
\affiliation{OzGrav, School of Physics \& Astronomy, Monash University, Clayton 3800, Victoria, Australia}
\author[0000-0002-4559-8427]{N.~Adhikari}
\affiliation{University of Wisconsin-Milwaukee, Milwaukee, WI 53201, USA}
\author[0000-0002-5731-5076]{R.~X.~Adhikari}
\affiliation{LIGO Laboratory, California Institute of Technology, Pasadena, CA 91125, USA}
\author{V.~K.~Adkins}
\affiliation{Louisiana State University, Baton Rouge, LA 70803, USA}
\author{V.~B.~Adya}
\affiliation{OzGrav, Australian National University, Canberra, Australian Capital Territory 0200, Australia}
\author{C.~Affeldt}
\affiliation{Max Planck Institute for Gravitational Physics (Albert Einstein Institute), D-30167 Hannover, Germany}
\affiliation{Leibniz Universit\"at Hannover, D-30167 Hannover, Germany}
\author{D.~Agarwal}
\affiliation{Inter-University Centre for Astronomy and Astrophysics, Pune 411007, India}
% ...

\collaboration{3000}{The LIGO Scientific Collaboration, the Virgo Collaboration, and the KAGRA Collaboration}
  }
{
  \author{The LIGO Scientific Collaboration}
  \affiliation{LSC}
  \author{The Virgo Collaboration}
  \affiliation{Virgo}
  \author{The KAGRA Collaboration}
  \affiliation{KAGRA}
}
\date{\today}

\begin{abstract}
We use \Nevgwcosmo gravitational-wave sources from the Third LIGO--Virgo--KAGRA Gravitational-Wave Transient Catalog (GWTC--3) to estimate the Hubble parameter $H(z)$, including its current value, the Hubble constant $H_0$.
Each gravitational-wave (GW) signal provides the luminosity distance to the source and we estimate the corresponding redshift using two methods: the redshifted masses and a galaxy catalog. Using the binary black hole (BBH) redshifted masses, we simultaneously infer the source mass distribution and $H(z)$.
The source mass distribution displays a peak around $34\, \Msol$, followed by a drop-off.
Assuming this mass scale does not evolve with redshift results in a $H(z)$ measurement, yielding $H_0=\HnotnoticarogwwithBNSlogPLG \hu$ ($68\%$ credible interval) when combined with the $H_0$ measurement from GW170817 and its electromagnetic counterpart. This represents an improvement of \HnoticarogwimproGWTConePLG\%  with respect to the $H_0$ estimate from GWTC--1. 
The second method associates each GW event with its probable host galaxy in the catalog \gladeplus, statistically marginalizing over the redshifts of each event's potential hosts.
Assuming a fixed BBH population, we estimate a value of $H_0=\HnotgwcosmoBNSKband \hu$ with the galaxy catalog method, an improvement of \HnotgwcosmoimproGWTConeKband\% with respect to our GWTC--1 result and 20\% with respect to recent $H_0$ studies using GWTC--2 events.
However, we show that this result is strongly impacted by assumptions about the BBH source mass distribution; the only event which is not strongly impacted by such assumptions (and is thus informative about $H_0$) is the well-localized event GW190814.
\end{abstract}

\keywords{gravitational waves, cosmology: observations, cosmological parameters}

\section{Introduction}\label{sec:intro}

The discovery of a gravitational wave (GW) signal from a binary neutron star (BNS) merger~\citep{TheLIGOScientific:2017qsa} and the kilonova emission from its remnant~\citep{2017Sci...358.1556C,LIGOScientific:2017ync} provided the first GW standard siren measurement of the cosmic expansion history~\citep{Abbott:2017xzu}.
As pointed out by \citet{1986Natur.323..310S}, the GW signal from a compact binary coalescence directly measures the luminosity distance to the source without any additional distance calibrator, earning these sources the name ``standard sirens"~\citep{Holz:2005df}. 
Measuring the cosmic expansion as a function of cosmological redshift is one of the key avenues with which to explore the constituents of the Universe, along with the other canonical probes such as the cosmic microwave background (CMB; \citealp{Spergel:2003cb, Spergel:2006hy,2011ApJS..192...18K, Ade:2013zuv, Planck:2015fie, Aghanim:2018eyx}), baryon acoustic oscillations \citep{Eisenstein:1997ik,Eisenstein:1997jh,2013AJ....145...10D,Alam:2016hwk}, type Ia supernovae \citep{Riess:1996pa, Perlmutter:1998np, Riess:2016jrr, Freedman:2017yms, Riess:2019cxk}, strong gravitational lensing \citep{2010ApJ...711..201S, Wong:2019kwg}, and cosmic chronometers \citep{Jimenez:2019onw}. 

Even though GW sources are excellent distance tracers, using them to study the expansion history also requires measurement of their redshift. The redshift information is usually degenerate with the source masses in the GW signal, as the redshifted masses affect the GW frequency evolution. However, several techniques are proposed to infer the redshift of GW sources and break the mass-redshift degeneracy. For sources with confirmed electromagnetic counterparts, the host galaxy and its redshift can be determined directly~\citep{Holz:2005df,Dalal:2006qt, PhysRevD.77.043512, 2010ApJ...725..496N, Abbott:2017xzu,Chen:2017rfc,Feeney:2018mkj}. For sources without an electromagnetic counterpart, alternative techniques to infer the source redshift include comparing the redshifted mass distribution to an astrophysically-motivated source mass distribution~\citep{1993ApJ...411L...5C,Taylor:2012db,Farr:2019twy,You:2020wju,2021PhRvD.104f2009M}, obtaining statistical redshift information from galaxy catalogs~\citep{1986Natur.323..310S, PhysRevD.77.043512, 2012PhRvD..86d3011D,Nishizawa:2016ood, Chen:2017rfc,2018PhRvD..98b3502N,2019ApJ...871L..13F, 2019ApJ...876L...7S, Gray:2019ksv, Yu:2020vyy,  2020ApJ...900L..33P,Borhanian:2020vyr,Finke:2021aom}, comparing the spatial clustering between GW sources and galaxies \citep{Oguri:2016dgk,Mukherjee:2019wcg,Bera:2020jhx,2021PhRvD.103d3520M}, leveraging external knowledge of the source redshift distribution~\citep{Ding:2018zrk,Ye:2021klk}, and exploiting the tidal distortions of neutron stars~\citep{2012PhRvL.108i1101M, Chatterjee:2021xrm}. 

The third LIGO--Virgo--KAGRA GW transient catalog (GWTC--3) \citep{O3bCat_PLACEHOLDER} contains 90 compact binary coalescence candidate events with at least a 50\% probability of being astrophysical in origin. 
Out of the events from the third observing run, a notable electromagnetic counterpart has been claimed only for the high-mass binary black hole (BBH) event GW190521~\citep{2020PhRvL.124y1102G}. However, the significance of this association has been re-assessed by several authors, who found insufficient evidence to claim a confident association \citep{Ashton:2020kyr,DePaolis:2020onl,Palmese:2021wcv}. In this work, we do not include the redshift information from this putative electromagnetic counterpart signal. The only standard siren with an electromagnetic counterpart in GWTC--3 remains the BNS GW170817. 

The remainder of this paper is organized as follows. In Sec.~\ref{sec:method} we discuss the statistical methods used to infer the cosmological parameters with and without galaxy catalog information. In Sec.~\ref{sec:events} we discuss the properties of the GW events and galaxy catalogs used in this paper. In Sec.~\ref{sec:results} we present the results of our analyses and in Sec.~\ref{sec:discussion} we discuss their implications for the cosmological parameters. Finally, in Sec.~\ref{sec:conclusions} we present our conclusions.

\section{Method \label{sec:method}}

We use two analysis methods: (i) jointly fitting the cosmological parameters and the source population properties of BBHs, without using galaxy catalog information (\citealt{2021PhRvD.104f2009M}; Sec.~\ref{sec:icarogw}), and (ii) fixing the source population properties, and inferring the cosmological parameters using statistical galaxy catalog information (\citealt{Gray:2019ksv}; Sec.~\ref{sec:gwcosmo_method}).

\subsection{Hierarchical inference without galaxy surveys \label{sec:icarogw}}
The GW event catalog can be described by two sets of parameters: a set of \textit{population hyper-parameters} $\Phi$ that are common to the entire population of GW sources, and a set of \textit{intrinsic parameters} that are unique for each event. The cosmological population hyper-parameters in this work are the cosmological parameters for a flat Universe. For the redshift range considered in the analysis, the contribution to the total energy density from radiation and neutrinos is negligible. Hence, we consider the dark energy density today as $\Omega_{\rm DE}(z)=1-\Omega_{\rm m}$. The cosmological parameters considered are, therefore, the Hubble constant $H_0$, the matter density $\Omega_{\rm m}$ and dark energy equation of state (EOS) parameter $w(z)=w_0$. \citep{Chevallier:2000qy}. The dark energy EOS is defined by $w=p/\rho$, where for the standard $\Lambda \text{CDM}$ we have $w=-1$. Additionally, the set of hyper-parameters $\Phi$ includes parameters describing the source mass distribution and the merger rate density as a function of redshift. 

Given a set of $N_{\rm obs}$ GW detections associated with the data $\{x\}=(x_1,...,x_{\rm N_{obs}})$, the posterior on $\Phi$ can be expressed as \citep{Mandel:2018mve,2019PASA...36...10T,Vitale:2020aaz}

\begin{equation}
    p(\Phi|\{x\},N_{\rm obs}) = p(\Phi) \prod_{i=1}^{N_{\rm obs}} \frac{\int p(x_i|\Phi,\theta)p_{\rm pop}(\theta|\Phi)\text{d}\theta}{\int p_{\rm det}(\theta,\Phi)p_{\rm pop}(\theta|\Phi)\text{d}\theta},
    \label{eq:post}
\end{equation}
where $p(\Phi)$ is a prior on the population parameters, $\theta$ is the set of parameters intrinsic to each GW event, such as spins, masses, redshift etc, $p(x_i|\Phi,\theta)$ is the GW likelihood, $p_{\rm det}(\theta,\Phi)$ is the probability of detecting a GW event with intrinsic parameters $\theta$ and for population hyper-parameters $\Phi$, and $p_{\rm pop}(\theta|\Phi)$ is a \textit{population modelled} prior.
The denominator in Eq.~\eqref{eq:post} correctly normalizes  the posterior and takes into account \textit{selection effects\/} \citep{Mandel:2018mve}.
We use the hierarchical statistical framework to infer the population parameters $\Phi$ and the prior that they induce on the distributions of the GW parameters.

The intrinsic GW parameters that are interesting for cosmology are those which provide information about the redshift $z$ of the source.
For sources which are detected at cosmological distances, GWs provide a measurement of the \textit{redshifted} masses $m_{1}^{\mathrm{det}},m_{2}^\mathrm{det}$ and luminosity distance $D_\mathrm{L}$, rather than the redshift and \textit{source} masses $m_{1},m_{2}$, where

\begin{equation}
m_{i}=\frac{m_{i}^\mathrm{det}}{1+z(D_\mathrm{L};H_0,\Omega_{\rm m},w_0)}.
\label{eq:masses}
\end{equation}

The relation between source mass and redshifted mass can then be used to probe cosmology even in the absence of an explicit electromagnetic (EM) counterpart \citep{Taylor:2011fs, Taylor:2012db} provided source mass scale can be well characterized. This approach is more effective if the source mass distribution displays sharp features \citep{Farr:2019twy, Ezquiaga:2020tns, You:2020wju,2021PhRvD.104f2009M}.

\subsubsection{Population models}

We model the BBH population since BBHs represent the majority of the detected sources.  We now give a general overview of the source mass and redshift models used in this paper; see App.~\ref{app:priors} for a complete description of the population models.

We describe the underlying distribution in redshift and source masses as

\begin{eqnarray}
 p_{\rm pop}(\theta|\Phi_m,H_0,\Omega_{\rm m},w_0) &=& C \: p(m_{1},m_{2}|\Phi_m) \psi(z|\gamma,k,z_\text{p}) \nonumber \\ && \times   \frac{p(z|H_0,w_0,\Omega_{\rm m})}{1+z},
 \label{eq:popinduced}
\end{eqnarray}
where $C$ is a normalization factor, $p(m_1,m_2|\Phi_m)$ is the source frame mass  distribution, $\Phi_m$ refers to all the  population parameters not related to cosmology, the $(1+z)$ term encodes the clock difference between the source frame and detector frame, and $p(z|H_0,w_0,\Omega_{\rm m})$ is the redshift prior which is taken to be uniform in comoving volume. The term $\psi(z|\gamma,k,z_\text{p})$ describes the redshift evolution of the merger rate with a parameterization similar to that of \citet{2014ARA&A..52..415M}, characterized by a low-redshift power-law slope $\gamma$, a peak at redshift $z_\text{p}$, and a high-redshift power-law slope $k$ after the peak, or

\begin{equation}
   \psi(z|\gamma,k,z_\text{p}) = [1+(1+z_\text{p})^{-\gamma - k}]  \frac{(1+z)^{\gamma}}{1+\left[(1+z)/(1+z_\text{p}) \right]^{\gamma + k}}.
\end{equation}

The above rate evolution model is more complex than that in \citet{Fishbach:2018edt,Abbott:2020gyp,O3bPops_PLACEHOLDER}; this is because when we vary the cosmological parameters, the GW observations may be pushed to higher redshift $z>2$, past the peak of the star formation rate \citep{2014ARA&A..52..415M}. 

The source mass models are factorized as
\begin{equation}
 p(m_{1},m_{2}|\Phi_m)=p(m_{1}|\Phi_m)p(m_{2}|m_{1},\Phi_m),   
\end{equation}
where the secondary mass is modeled by a power-law distribution between a minimum mass $\mmin$ and maximum mass $m_{1}$. For the primary mass, we implement three phenomenological mass models used in \citet{LIGOScientific:2018jsj, Abbott:2020gyp}.

The first phenomenological model, is the \PL model, describes the mass distribution as a power law (PL) between a minimum mass $\mmin$ and a maximum mass $\mmax$~\citep{2017ApJ...851L..25F}. The BBH mass distribution inferred from GWTC--2 was more complicated than a \PL PL, and the second and third models are extensions of the \PL model that contain more complex structures to better fit the mass distribution~\citep{Abbott:2020gyp}. 
The second model, \BPL, consists of two PLs attached at a break--point~\citep{Abbott:2020gyp}.
The third model is a superposition of a \PL and a Gaussian component referred to as \PLG, in which the primary mass distribution is described by a PL with the addition of a Gaussian peak with mean $\mu_\text{g}$ and variance $\sigma_\text{g}^2$~\citep{Talbot:2018cva}. Using the \BPL and \PLG models, GWTC--2 revealed an excess of BBH systems with primary masses in the range $\sim30$--$40\,\Msol$, followed by a drop-off in the merger rate at high masses~\citep{Abbott:2020gyp}. This structure in the PL, modeled either as a break or a Gaussian peak, may represent the imprint of (pulsational) pair-instability supernovae~\citep{Fryer:2000my,Heger:2001cd,2019ApJ...887...53F,Renzo:2020lwl,Umeda:2020fih}. 


In a companion paper investigating the GWTC--3 population \citep{O3bPops_PLACEHOLDER} we show the evidence for sub-structures in the BHs primary mass spectrum around $\sim 10 \Msol$. However we also find that the simpler \PLG model is still one of the models preferred by the GWTC--3 data. For this reason, and for simplicity in this paper, we only adopt models which are characterized by a single structure (\BPL and \PLG) to describe the excess of BHs observed around $35 \Msol$; this also corresponds to the binaries that we can observe at higher redshifts and for which source mass assumptions could be important.

In order to infer $H(z)$ from the BBH population, a crucial assumption is that the source mass distribution is independent of redshift~\citep{2021ApJ...912...98F}. In most BBH formation scenarios, we expect some evolution of the mass distribution with redshift, due to factors such as the metallicity evolution of the Universe~\citep{2000ARA&A..38..613K,2010ApJ...715L.138B} and the dependence of the delay time between BBH formation and merger on BBH properties~\citep{2016MNRAS.462..844K,2021arXiv210705702G,2021arXiv211001634V}. Nevertheless, if mass features, such as the break in the \BPL model or the peak in the \PLG model, are caused by the pair-instability supernova process, their location is thought to stay constant within a few solar masses across cosmic time~\citep{2019ApJ...887...53F}. The presence of these sharp mass features drives our cosmological constraints~\citep{Farr:2019twy,2021PhRvD.104f2009M}. Moreover, the BBH mass distribution is expected to evolve only weakly over the range of redshift accessible to current observations, at a level below current statistical uncertainties~\citep{2021ApJ...914L..30F,2021arXiv211001634V}. Although the BH mass spectrum at \emph{formation} may vary with cosmic time, BBH channels typically predict a wide distribution of delay times between formation and merger, which tends to wash out any dependence of BBH mass on merger redshift~\citep{2019MNRAS.487....2M}. 

In the following we will neglect the selection effect of spin distribution as the detection probability due to their inclusion should not vary by more than a factor of two \citep{PhysRevD.98.083007}. This is indeed a negligible term with respect to the statistical uncertainties on our posteriors (see Sec. \ref{sec:discussion}) and the dependence of the selection bias with respect to other parameters such as $H_0$ and $\gamma$, for which it nearly follows a power-law. 

\subsection{Statistical galaxy catalog method \label{sec:gwcosmo_method}}

We also use the \texttt{gwcosmo}  code \citep{Gray:2019ksv} in the pixelated sky scheme \citep{2021arXiv211104629G}, i.e.\ using the HEALPix pixelization algorithm \citep{2005ApJ...622..759G, Zonca2019}, to infer $H_0$ using information from galaxy surveys. This method assumes a fixed source mass distribution, as well as a fixed-rate evolution for the binaries, and estimates $H_0$ from the GW data using galaxy catalogs to provide statistical information about the GW source redshifts. 

When including galaxy catalog information, the prior on redshift can be replaced by the distribution of galaxies in the survey. However, Eq.~\eqref{eq:post} needs to be modified in order to take into account \textit{completeness corrections}. These extra terms account for the impact of incompleteness, i.e.\ missing galaxies, due to the limited sensitivity of the catalog, in the GW localization volume.
In this case, the posterior is given by
\begin{eqnarray}
     & p(H_0|x, N_{\rm obs},\Phi_m) = p(H_0) p(N_{\rm obs}|H_0,\Phi_m) \times \nonumber \\& \prod\limits_{i=1}^{N_{\rm obs}}\sum\limits_{g \in[G,\bar{G}]} p(x_i|\hat{d},H_0,\Phi_m,g)p(g|H_0,\Phi_m,\hat{d}),
\end{eqnarray}
where $G$ is the hypothesis that the GW host galaxy is included in the catalog and $\bar{G}$ that it is not and $p(g|H_0,\Phi_m,\hat{d})$ expresses their probabilities for $g \in [G,\bar{G}]$. The term $p(N_{\rm obs}|H_0,\Phi_{m})$ is the probability of having $N_{\rm obs}$ detections. We analytically marginalize over this by assuming a uniform in log rate prior. The notation $\hat{d}$ indicates the hypothesis that an event has been detected. The likelihoods $p(x_i|\hat{d},H_0,\Phi_m,g)$ are built from the GW data and corrected for the selection effects in the case that the host galaxy is, and is not, inside the catalogue; see \citet{Gray:2019ksv}.

We implement an improved version of the analysis presented in \citet{Abbott:2019yzh} that can estimate $H_0$ for any given sky direction covered by the GW localization by dividing the sky into equal-area pixels. In each pixel, the apparent magnitude threshold ($m_{\rm thr}$) is taken to be the median of the apparent magnitudes of all the galaxies inside that pixel. This assumption is a conservative choice for approximating the impact of catalog completeness: all galaxies with apparent magnitude fainter than the defined threshold are excluded from the analysis. Using this $m_{\rm thr}$ the completeness is assessed and the $H_0$ likelihoods are calculated in each pixel. In the end, all the pixel likelihoods are combined using weights proportional to the GW posterior probability in each pixel to give the final $H_0$ posterior of each GW event. Pixels with no GW support make zero contribution, so only the pixels within the 99.9$\%$ sky area are used. 

\begin{figure*}[htp!]
    \centering
    \includegraphics[scale=0.9]{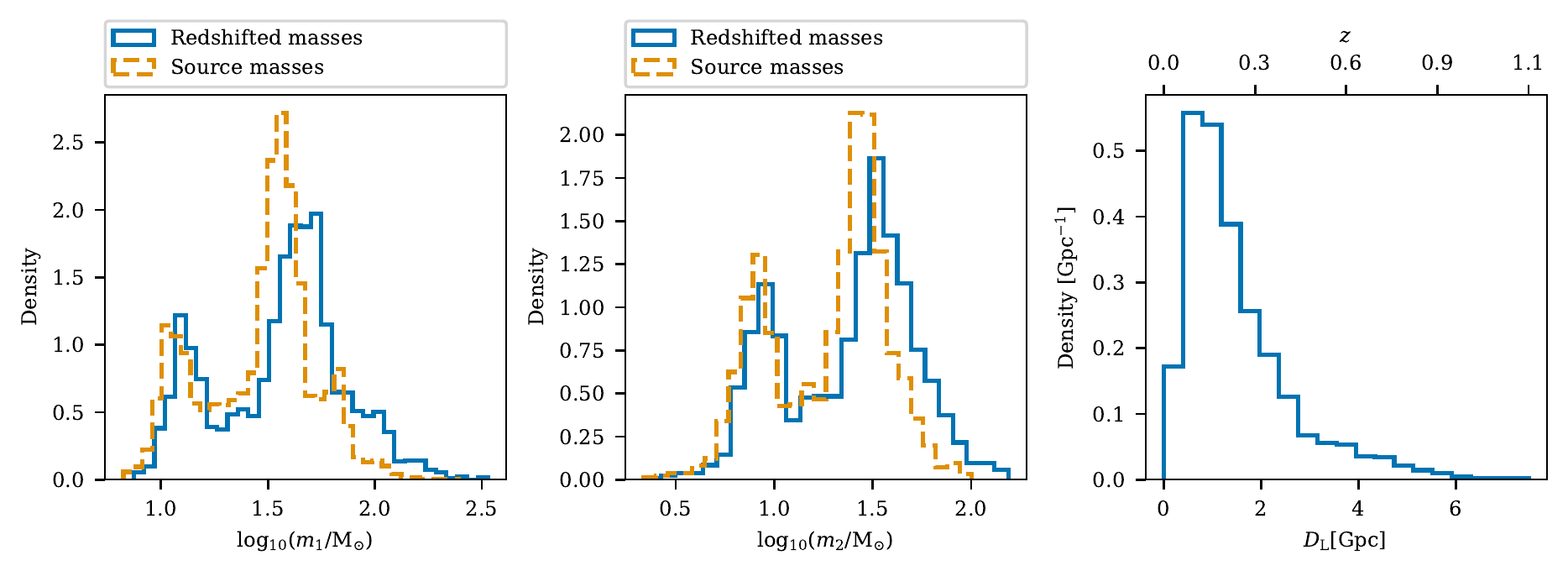}
    \caption{Distribution of the mass and luminosity distance parameters for the \Nevicarogw BBH events with SNR $>  11$. The figure is generated using a Planck cosmology with $H_0=67.9 \hu$ and $\Omega_{\rm m}=0.3065$ with a $D^2_L$ prior and a uniform prior on the detector frame masses ad stacking posterior samples for each event. This figure is only representative of the events reported in Tab.~\ref{tab:recap} and does not indicate the population reconstruction. \textit{Left:} Distribution of the  primary detector frame masses (blue solid line) and source frame masses (orange dashed line). \textit{Middle:} Same but for the secondary source mass. \textit{Right:} Distribution of the luminosity distance (bottom axis) and redshift (top axis).}
    \label{fig:representation_tab1}
\end{figure*}

These improvements are necessary to correct for galaxy catalog incompleteness in the case that the galaxy surveys contained within the catalog are less sensitive in particular sky areas, such as the directions of the galactic plane. Moreover, the analysis can take into account the fact that the GW luminosity distance posterior conditioned on the sky position might significantly change between different sky positions. The combination of galaxy redshift information and luminosity distance estimation change from pixel to pixel, leading to a more robust estimation of $H_0$.

\section{Events and catalogs selection \label{sec:events}}

\subsection{GW events}

For our main result, we select \Nevgwcosmo GW events with network matched filter signal-to-noise ratio (SNR) $>11$ and Inverse False Alarm Rate (IFAR) higher than $4~{\rm yr}$, taking their maximum across the different search pipelines from GWTC--3 \citep{O3bCat_PLACEHOLDER}, and no plausible instrumental origin.

We also consider events identified by different SNR choices to explore possible systematics in the computation of selection effects, see App.~\ref{app:fullicarogw}. In the remainder of the paper we will shortly refer to these different ensembles by quoting the threshold SNR choices.
Of the \Nevgwcosmo events with SNR $>11$, \Nevicarogw\, are BBH detections, 2 are the BNS events GW170817 \citep{TheLIGOScientific:2017qsa} and GW190425 \citep{Abbott:2020uma}, 2 are the NSBH events GW200105 and GW200115 \citep{2021ApJ...915L...5A} and one is the asymmetric mass binary GW190814 \citep{Abbott:2020khf}. A visual representation of the population of BBHs that we detected is provided in Fig.~\ref{fig:representation_tab1}, where we show the distribution of detector frame masses and luminosity distance of the BBHs. 
We have tabulated all the GW sources used in this analysis in Table \ref{tab:search_setup_parameter}, mentioning their source properties, sky localization error, the 3D localization volume, number of galaxies in the catalog within the localization volume and the probability that the GW host is present in the \gladeplus catalog \citep{Dalya:2018cnd, Dalya:2021ewn}. Note that differently from \citep{O3bCat_PLACEHOLDER}, the estimation of masses and distances are reported using a prior $\propto D_L^2$ and not uniform in comoving volume since we are interested to show these values using cosmology-agnostic priors.
For the events detected during O1, O2 and O3 we use combined posterior samples from the \texttt{IMRPhenom}  \citep{Thompson:2020nei,Pratten:2020ceb} and \texttt{SEOBNR}  \citep{Ossokine:2020kjp,Matas:2020wab} families, while for the two NSBH events GW200105 and GW200115 we use posterior samples generated with low spin priors \citep{2021ApJ...915L...5A}.\footnote{Events in the analysis showing differences in posterior samples with different waveforms are GW191109\_010717 and GW200129\_065458 \citep{O3bCat_PLACEHOLDER}. These differences are mostly related to the effective and precession spin parameters which are not considered in this analysis.}


\begin{longrotatetable}[!p]
 \begin{deluxetable*}{lcccccccccccc}
 \tabletypesize{\tiny}
\tablecaption{
\label{tab:search_setup_parameter}
\scriptsize{This table reports all of the GW events considered in this paper and summarizes some of their properties reported with their median values and 90\% symmetric credible intervals. First, second and third columns: GW event label, detected SNR (highest among the different pipelines, see \citep{O3bCat_PLACEHOLDER}) and inverse false alarm rate. Fourth, fifth, sixth columns: estimated primary and secondary detector frame masses, luminosity distance. Seventh, eighth and ninth columns: primary and secondary source frame masses, and redshift assuming a reference cosmology.  Tenth and eleventh columns: sky localization area, and 3D localization comoving volume. The twelfth column lists the number of galaxies in \gladeplus inside the localization volume for each event with observed K--band ($B_J$--band in parenthesis), while the thirteenth column reports the probability $p(G|z, H_0)$ that \gladeplus detected the GW event host galaxy using the K--band ($B_J$ in parenthesis) luminosity function calculated for a fiducial flat $\Lambda$CDM cosmology with $H_0=67.9\, \hu$ and $\Omega_{\rm m}=0.3065$. The lower and upper bounds on these probabilities are derived from the completeness fractions at the boundaries of the 90\% localization volume. We report these values using a distance prior proportional to $D_L^2$ and a uniform in detector frame masses prior. The events in bold are the ones with SNR$>11$ entering all the analyses while the others are the ones used only for the systematic studies in App.~\ref{app:fullicarogw}.}
}
\tablehead{\colhead{Name} & \colhead{SNR} & \colhead{IFAR} & \colhead{$m_{1}^{\mathrm{det}}$} & \colhead{$m_{2}^{\mathrm{det}}$} & \colhead{$D_\mathrm{L}$} & \colhead{$m_{1}$} & \colhead{$m_{2}$} & \colhead{$z$} & \colhead{$\Delta \Omega$} & \colhead{$\Delta V_c$} & \colhead{$N_{\rm gal}$} & \colhead{$p(G|z,H_0)$} \\
\colhead{} & \colhead{} & \colhead{${\rm [yr]}$} & \colhead{${\rm [\Msol]}$} & \colhead{${\rm [\Msol]}$} & \colhead{$[{\rm Mpc}]$} & \colhead{${\rm [\Msol]}$} & \colhead{${\rm [\Msol]}$} & \colhead{} & \colhead{${\rm [deg^2]}$} & \colhead{${\rm [Gpc^3]}$} & \colhead{} & \colhead{}
}
\startdata
\textbf{GW150914} & 24.4 & $1 \times 10^{7}$ & $39^{+5}_{-3}$ & $33^{+3}_{-5}$ & $400^{+100}_{-200}$ & $35^{+5}_{-3}$ & $30^{+3}_{-4}$ & $0.09^{+0.03}_{-0.03}$ & 200 & $3 \times 10^{-4}$ & 3000($3 \times 10^{4}$) & 50-5(99-72)\\ 
GW151012 & 10.0 & 100 & $27^{+16}_{-6}$ & $17^{+4}_{-6}$ & $1100^{+500}_{-500}$ & $23^{+14}_{-5}$ & $14^{+4}_{-5}$ & $0.21^{+0.09}_{-0.09}$ & 2000 & 0.05 & 6000($5 \times 10^{5}$) & 5-0(73-10) \\ 
\textbf{GW151226} & 13.1 & $1 \times 10^{7}$ & $15^{+8}_{-4}$ & $8^{+2}_{-2}$ & $400^{+200}_{-200}$ & $14^{+8}_{-3}$ & $8^{+2}_{-2}$ & $0.09^{+0.04}_{-0.04}$ & 1000 & 0.003 & $1 \times 10^{4}$($1 \times 10^{5}$) & 58-4(100-70)\\ 
\textbf{GW170104} & 13.0 & $1 \times 10^{7}$ & $37^{+8}_{-6}$ & $24^{+6}_{-6}$ & $1000^{+400}_{-400}$ & $31^{+7}_{-6}$ & $20^{+5}_{-5}$ & $0.20^{+0.07}_{-0.08}$ & 900 & 0.02 & 8000($4 \times 10^{5}$) & 8-0(76-17)\\ 
\textbf{GW170608} & 15.4 & $1 \times 10^{7}$ & $12^{+6}_{-2}$ & $8^{+1}_{-2}$ & $300^{+100}_{-100}$ & $11^{+5}_{-2}$ & $8^{+1}_{-2}$ & $0.07^{+0.02}_{-0.02}$ & 400 & $4 \times 10^{-4}$ & 4000($2 \times 10^{4}$) & 69-19(100-87)\\ 
GW170729 & 10.8 & 6 & $80^{+20}_{-10}$ & $50^{+10}_{-20}$ & $3000^{+1000}_{-1000}$ & $50^{+20}_{-10}$ & $32^{+10}_{-9}$ & $0.5^{+0.2}_{-0.2}$ & 1000 & 0.3 & 60($8 \times 10^{4}$) & 0-0(12-0) \\ 
\textbf{GW170809} & 12.4 & $1 \times 10^{7}$ & $42^{+10}_{-7}$ & $28^{+6}_{-6}$ & $1000^{+300}_{-400}$ & $35^{+9}_{-6}$ & $24^{+5}_{-5}$ & $0.20^{+0.05}_{-0.07}$ & 300 & 0.004 & 2000($1 \times 10^{5}$) & 4-0(70-21)\\ 
\textbf{GW170814} & 16.3 & $1 \times 10^{7}$ & $34^{+6}_{-3}$ & $29^{+3}_{-5}$ & $600^{+100}_{-200}$ & $30^{+6}_{-3}$ & $25^{+3}_{-4}$ & $0.12^{+0.03}_{-0.04}$ & 90 & $2 \times 10^{-4}$ & 1000($1 \times 10^{4}$) & 27-1(91-60)\\ 
\textbf{GW170817} & 33.0 & $1 \times 10^{7}$ & $1.48^{+0.15}_{-0.09}$ & $1.28^{+0.09}_{-0.12}$ & $40^{+7}_{-15}$ & $1.47^{+0.15}_{-0.09}$ & $1.27^{+0.09}_{-0.11}$ & $0.009^{+0.002}_{-0.003}$ & 20 & $2 \times 10^{-8}$ & 10(6) & 100-100(100-100)\\ 
\textbf{GW170818} & 11.3 & $2 \times 10^{4}$ & $43^{+9}_{-5}$ & $32^{+5}_{-7}$ & $1000^{+400}_{-300}$ & $35^{+7}_{-5}$ & $27^{+4}_{-5}$ & $0.21^{+0.07}_{-0.06}$ & 40 & $6 \times 10^{-4}$ & 80($1 \times 10^{4}$) & 2-0(55-9)\\ 
\textbf{GW170823} & 11.5 & $1 \times 10^{7}$ & $53^{+13}_{-8}$ & $39^{+8}_{-10}$ & $1900^{+800}_{-900}$ & $39^{+11}_{-7}$ & $29^{+6}_{-7}$ & $0.4^{+0.1}_{-0.1}$ & 2000 & 0.2 & 700($3 \times 10^{5}$) & 0-0(35-0)\\ 
\textbf{GW190408\textunderscore181802} & 14.8 & $1 \times 10^{5}$ & $32^{+6}_{-4}$ & $24^{+4}_{-5}$ & $1600^{+400}_{-600}$ & $24^{+5}_{-3}$ & $18^{+3}_{-4}$ & $0.30^{+0.06}_{-0.10}$ & 100 & 0.003 & 80($2 \times 10^{4}$) & 0-0(39-3)\\ 
\textbf{GW190412} & 19.7 & $1 \times 10^{5}$ & $35^{+5}_{-6}$ & $10^{+2}_{-1}$ & $700^{+100}_{-200}$ & $30^{+5}_{-5}$ & $8.3^{+1.6}_{-0.9}$ & $0.15^{+0.03}_{-0.03}$ & 20 & $3 \times 10^{-5}$ & 100(4000) & 7-0(76-51)\\ 
GW190413\textunderscore134308 & 10.3 & 6 & $80^{+20}_{-10}$ & $60^{+20}_{-20}$ & $5000^{+2000}_{-2000}$ & $45^{+13}_{-10}$ & $31^{+10}_{-10}$ & $0.8^{+0.3}_{-0.3}$ & 500 & 0.2 & $0$(70) & 0-0(0-0) \\ 
GW190421\textunderscore213856 & 10.5 & 400 & $62^{+15}_{-9}$ & $48^{+10}_{-13}$ & $3000^{+1000}_{-1000}$ & $41^{+10}_{-7}$ & $31^{+7}_{-8}$ & $0.5^{+0.2}_{-0.2}$ & 1000 & 0.2 & 1($2 \times 10^{4}$) & 0-0(5-0) \\ 
\textbf{GW190425} & 12.9 & 30 & $2.1^{+0.6}_{-0.4}$ & $1.4^{+0.3}_{-0.3}$ & $160^{+70}_{-70}$ & $2.0^{+0.6}_{-0.3}$ & $1.4^{+0.3}_{-0.3}$ & $0.03^{+0.01}_{-0.02}$ & 8000 & 0.002 & $3 \times 10^{4}$($6 \times 10^{4}$) & 95-63(100-100)\\ 
\textbf{GW190503\textunderscore185404} & 12.8 & $1 \times 10^{5}$ & $55^{+11}_{-10}$ & $40^{+10}_{-10}$ & $1500^{+700}_{-700}$ & $43^{+9}_{-8}$ & $29^{+7}_{-8}$ & $0.3^{+0.1}_{-0.1}$ & 90 & 0.005 & 100($4 \times 10^{4}$) & 0-0(46-1)\\ 
\textbf{GW190512\textunderscore180714} & 12.4 & $1 \times 10^{5}$ & $29^{+7}_{-7}$ & $16^{+5}_{-3}$ & $1500^{+500}_{-600}$ & $23^{+5}_{-6}$ & $13^{+4}_{-2}$ & $0.29^{+0.08}_{-0.10}$ & 200 & 0.008 & 100($3 \times 10^{4}$) & 0-0(41-2)\\ 
\textbf{GW190513\textunderscore205428} & 12.9 & $8 \times 10^{4}$ & $50^{+10}_{-10}$ & $25^{+10}_{-6}$ & $2200^{+900}_{-800}$ & $35^{+10}_{-9}$ & $18^{+7}_{-4}$ & $0.4^{+0.1}_{-0.1}$ & 500 & 0.04 & 20($3 \times 10^{4}$) & 0-0(18-0)\\ 
\textbf{GW190517\textunderscore055101} & 11.3 & 3000 & $51^{+15}_{-8}$ & $35^{+8}_{-10}$ & $2000^{+2000}_{-1000}$ & $36^{+12}_{-8}$ & $25^{+7}_{-7}$ & $0.4^{+0.3}_{-0.2}$ & 400 & 0.1 & 90($7 \times 10^{4}$) & 0-0(28-0)\\ 
\textbf{GW190519\textunderscore153544} & 13.9 & $1 \times 10^{5}$ & $100^{+20}_{-10}$ & $60^{+20}_{-20}$ & $3000^{+2000}_{-1000}$ & $60^{+10}_{-10}$ & $40^{+10}_{-10}$ & $0.5^{+0.3}_{-0.2}$ & 700 & 0.2 & 9($2 \times 10^{4}$) & 0-0(7-0)\\ 
\textbf{GW190521} & 14.4 & 5000 & $150^{+50}_{-20}$ & $120^{+30}_{-40}$ & $5000^{+2000}_{-2000}$ & $90^{+30}_{-20}$ & $70^{+20}_{-20}$ & $0.7^{+0.3}_{-0.3}$ & 900 & 0.4 & 4(900) & 0-0(0-0)\\ 
\textbf{GW190521\textunderscore074359} & 24.7 & $1 \times 10^{5}$ & $52^{+7}_{-6}$ & $41^{+6}_{-8}$ & $1300^{+400}_{-600}$ & $42^{+6}_{-5}$ & $33^{+5}_{-6}$ & $0.25^{+0.06}_{-0.10}$ & 500 & 0.01 & 700($3 \times 10^{4}$) & 1-0(59-8)\\ 
\textbf{GW190602\textunderscore175927} & 12.6 & $1 \times 10^{5}$ & $100^{+20}_{-20}$ & $70^{+20}_{-30}$ & $3000^{+2000}_{-1000}$ & $70^{+20}_{-10}$ & $50^{+10}_{-20}$ & $0.5^{+0.3}_{-0.2}$ & 700 & 0.2 & 2($2 \times 10^{4}$) & 0-0(6-0)\\ 
GW190620\textunderscore030421 & 10.9 & 90 & $80^{+20}_{-20}$ & $50^{+20}_{-20}$ & $3000^{+2000}_{-1000}$ & $60^{+20}_{-10}$ & $30^{+10}_{-10}$ & $0.5^{+0.2}_{-0.2}$ & 6000 & 2 & 200($1 \times 10^{5}$) & 0-0(6-0) \\ 
\textbf{GW190630\textunderscore185205} & 15.2 & $1 \times 10^{5}$ & $42^{+8}_{-7}$ & $28^{+6}_{-5}$ & $1000^{+500}_{-400}$ & $35^{+7}_{-6}$ & $23^{+5}_{-5}$ & $0.19^{+0.09}_{-0.08}$ & 1000 & 0.03 & 9000($5 \times 10^{5}$) & 8-0(76-12)\\ 
\textbf{GW190701\textunderscore203306} & 11.9 & 200 & $70^{+20}_{-10}$ & $60^{+10}_{-20}$ & $2200^{+800}_{-700}$ & $54^{+12}_{-8}$ & $41^{+8}_{-11}$ & $0.4^{+0.1}_{-0.1}$ & 40 & 0.003 & 9(6000) & 0-0(17-0)\\ 
\textbf{GW190706\textunderscore222641} & 12.7 & $2 \times 10^{4}$ & $110^{+20}_{-20}$ & $70^{+20}_{-30}$ & $5000^{+3000}_{-2000}$ & $60^{+20}_{-20}$ & $40^{+10}_{-10}$ & $0.8^{+0.3}_{-0.3}$ & 600 & 0.3 & $0$(10) & 0-0(0-0)\\ 
\textbf{GW190707\textunderscore093326} & 13.2 & $1 \times 10^{5}$ & $13^{+4}_{-2}$ & $10^{+1}_{-2}$ & $800^{+400}_{-400}$ & $12^{+3}_{-2}$ & $8^{+1}_{-2}$ & $0.17^{+0.06}_{-0.07}$ & 1000 & 0.02 & 6000($2 \times 10^{5}$) & 20-0(86-26)\\ 
\textbf{GW190708\textunderscore232457} & 13.1 & 3000 & $21^{+6}_{-2}$ & $15^{+2}_{-3}$ & $900^{+300}_{-400}$ & $17^{+5}_{-2}$ & $13^{+2}_{-3}$ & $0.18^{+0.06}_{-0.07}$ & $1 \times 10^{4}$ & 0.2 & $1 \times 10^{5}$($4 \times 10^{6}$) & 10-0(79-24)\\ 
\textbf{GW190720\textunderscore000836} & 11.6 & $1 \times 10^{5}$ & $16^{+7}_{-3}$ & $9^{+2}_{-3}$ & $800^{+700}_{-300}$ & $13^{+7}_{-3}$ & $8^{+2}_{-2}$ & $0.17^{+0.12}_{-0.06}$ & 500 & 0.02 & 5000($2 \times 10^{5}$) & 12-0(81-11)\\ 
\textbf{GW190727\textunderscore060333} & 12.1 & $1 \times 10^{5}$ & $59^{+13}_{-8}$ & $46^{+8}_{-13}$ & $4000^{+2000}_{-1000}$ & $37^{+9}_{-6}$ & $29^{+7}_{-8}$ & $0.6^{+0.2}_{-0.2}$ & 700 & 0.2 & $0$(5000) & 0-0(2-0)\\ 
\textbf{GW190728\textunderscore064510} & 13.4 & $1 \times 10^{5}$ & $14^{+8}_{-2}$ & $9^{+2}_{-3}$ & $900^{+200}_{-400}$ & $12^{+7}_{-2}$ & $8^{+2}_{-3}$ & $0.18^{+0.04}_{-0.07}$ & 400 & 0.004 & 2000($1 \times 10^{5}$) & 10-0(79-30)\\ 
\textbf{GW190814} & 22.2 & $1 \times 10^{5}$ & $24^{+1}_{-1}$ & $2.72^{+0.08}_{-0.09}$ & $240^{+40}_{-50}$ & $23^{+1}_{-1}$ & $2.59^{+0.08}_{-0.09}$ & $0.053^{+0.009}_{-0.010}$ & 20 & $9 \times 10^{-7}$ & 90(200) & 76-57(100-100)\\ 
\textbf{GW190828\textunderscore063405} & 16.6 & $1 \times 10^{5}$ & $44^{+7}_{-5}$ & $36^{+5}_{-6}$ & $2200^{+600}_{-900}$ & $32^{+6}_{-4}$ & $26^{+4}_{-5}$ & $0.40^{+0.09}_{-0.15}$ & 500 & 0.03 & 30($6 \times 10^{4}$) & 0-0(19-0)\\ 
\textbf{GW190828\textunderscore065509} & 11.1 & $3 \times 10^{4}$ & $31^{+9}_{-9}$ & $13^{+5}_{-3}$ & $1700^{+600}_{-600}$ & $24^{+7}_{-7}$ & $10^{+4}_{-2}$ & $0.31^{+0.10}_{-0.10}$ & 600 & 0.03 & 100($5 \times 10^{4}$) & 0-0(32-1)\\ 
\textbf{GW190910\textunderscore112807} & 13.4 & 300 & $57^{+9}_{-6}$ & $46^{+7}_{-9}$ & $1600^{+1100}_{-700}$ & $43^{+8}_{-6}$ & $35^{+6}_{-7}$ & $0.3^{+0.2}_{-0.1}$ & 9000 & 0.9 & 6000($1 \times 10^{6}$) & 0-0(41-0)\\ 
\textbf{GW190915\textunderscore235702} & 13.1 & $1 \times 10^{5}$ & $46^{+12}_{-8}$ & $32^{+7}_{-8}$ & $1700^{+700}_{-700}$ & $35^{+9}_{-6}$ & $24^{+5}_{-6}$ & $0.3^{+0.1}_{-0.1}$ & 300 & 0.02 & 700($1 \times 10^{5}$) & 0-0(35-1)\\ 
\textbf{GW190924\textunderscore021846} & 13.0 & $1 \times 10^{5}$ & $10^{+8}_{-2}$ & $6^{+1}_{-2}$ & $600^{+200}_{-200}$ & $9^{+7}_{-2}$ & $5^{+1}_{-2}$ & $0.12^{+0.04}_{-0.04}$ & 400 & 0.002 & 5000($6 \times 10^{4}$) & 34-1(93-55)\\ 
GW190929\textunderscore012149 & 10.3 & 6 & $100^{+30}_{-20}$ & $40^{+30}_{-20}$ & $4000^{+3000}_{-2000}$ & $60^{+20}_{-20}$ & $26^{+14}_{-10}$ & $0.6^{+0.4}_{-0.2}$ & 2000 & 1 & $0$($1 \times 10^{4}$) & 0-0(2-0) \\ 
GW190930\textunderscore133541 & 10.1 & 80 & $14^{+14}_{-3}$ & $9^{+2}_{-4}$ & $800^{+400}_{-300}$ & $12^{+12}_{-2}$ & $8^{+2}_{-3}$ & $0.16^{+0.07}_{-0.06}$ & 2000 & 0.02 & 7000($2 \times 10^{5}$) & 18-0(85-31) \\ 
GW191105\textunderscore143521 & 10.7 & 80 & $13^{+5}_{-2}$ & $9^{+2}_{-2}$ & $1200^{+400}_{-500}$ & $11^{+4}_{-2}$ & $7^{+1}_{-2}$ & $0.23^{+0.07}_{-0.08}$ & 800 & 0.02 & 3000($3 \times 10^{5}$) & 1-0(61-10) \\ 
\textbf{GW191109\textunderscore010717} & 15.8 & 6000 & $80^{+10}_{-8}$ & $60^{+20}_{-20}$ & $1300^{+1400}_{-700}$ & $60^{+10}_{-10}$ & $50^{+20}_{-10}$ & $0.3^{+0.2}_{-0.1}$ & 1000 & 0.3 & 5000($5 \times 10^{5}$) & 3-0(66-0)\\ 
GW191127\textunderscore050227 & 10.3 & 4 & $100^{+60}_{-30}$ & $40^{+30}_{-30}$ & $5000^{+3000}_{-3000}$ & $50^{+40}_{-20}$ & $20^{+20}_{-10}$ & $0.7^{+0.4}_{-0.4}$ & 1000 & 1 & 2(10000) & 0-0(3-0) \\ 
\textbf{GW191129\textunderscore134029} & 13.3 & $1 \times 10^{5}$ & $13^{+5}_{-3}$ & $8^{+2}_{-2}$ & $800^{+200}_{-300}$ & $11^{+4}_{-2}$ & $7^{+2}_{-2}$ & $0.16^{+0.04}_{-0.06}$ & 800 & 0.006 & 9000($2 \times 10^{5}$) & 13-0(82-35)\\ 
\textbf{GW191204\textunderscore171526} & 17.1 & $1 \times 10^{5}$ & $14^{+4}_{-2}$ & $9^{+2}_{-2}$ & $600^{+200}_{-200}$ & $12^{+3}_{-2}$ & $8^{+2}_{-2}$ & $0.13^{+0.03}_{-0.04}$ & 300 & 0.001 & 3000($5 \times 10^{4}$) & 21-0(87-52)\\ 
GW191215\textunderscore223052 & 10.9 & $1 \times 10^{5}$ & $33^{+9}_{-5}$ & $25^{+4}_{-5}$ & $2100^{+900}_{-900}$ & $24^{+7}_{-4}$ & $18^{+4}_{-4}$ & $0.4^{+0.1}_{-0.2}$ & 600 & 0.07 & 100($8 \times 10^{4}$) & 0-0(28-0) \\ 
\textbf{GW191216\textunderscore213338} & 18.6 & $1 \times 10^{5}$ & $14^{+4}_{-3}$ & $8^{+2}_{-2}$ & $300^{+100}_{-100}$ & $13^{+4}_{-3}$ & $7^{+2}_{-2}$ & $0.07^{+0.02}_{-0.03}$ & 500 & $4 \times 10^{-4}$ & 4000($3 \times 10^{4}$) & 67-19(100-86)\\ 
\textbf{GW191222\textunderscore033537} & 12.0 & $1 \times 10^{5}$ & $70^{+10}_{-10}$ & $50^{+10}_{-20}$ & $3000^{+2000}_{-2000}$ & $44^{+11}_{-8}$ & $33^{+9}_{-10}$ & $0.6^{+0.2}_{-0.3}$ & 2000 & 0.7 & 20($8 \times 10^{4}$) & 0-0(9-0)\\ 
GW191230\textunderscore180458 & 10.3 & 20 & $80^{+20}_{-10}$ & $60^{+10}_{-20}$ & $5000^{+2000}_{-2000}$ & $48^{+13}_{-9}$ & $36^{+10}_{-10}$ & $0.8^{+0.2}_{-0.3}$ & 800 & 0.2 & $0$(40) & 0-0(0-0) \\ 
\textbf{GW200105\textunderscore162426} & 13.9 & 5 & $9^{+3}_{-2}$ & $2.1^{+0.5}_{-0.4}$ & $300^{+100}_{-100}$ & $9^{+3}_{-2}$ & $1.9^{+0.5}_{-0.4}$ & $0.06^{+0.02}_{-0.03}$ & 7000 & 0.007 & $8 \times 10^{4}$($3 \times 10^{5}$) & 80-26(100-90)\\ 
\textbf{GW200112\textunderscore155838} & 17.6 & $1 \times 10^{5}$ & $45^{+8}_{-6}$ & $34^{+6}_{-7}$ & $1300^{+400}_{-500}$ & $36^{+7}_{-5}$ & $27^{+5}_{-6}$ & $0.25^{+0.07}_{-0.08}$ & 4000 & 0.09 & 8000($1 \times 10^{6}$) & 0-0(51-7)\\ 
\textbf{GW200115\textunderscore042309} & 11.5 & $1 \times 10^{5}$ & $7^{+2}_{-2}$ & $1.4^{+0.4}_{-0.2}$ & $300^{+200}_{-100}$ & $7^{+2}_{-2}$ & $1.3^{+0.4}_{-0.2}$ & $0.06^{+0.03}_{-0.03}$ & 700 & 0.001 & 7000($2 \times 10^{4}$) & 77-17(100-85)\\ 
GW200128\textunderscore022011 & 10.1 & 200 & $65^{+14}_{-9}$ & $50^{+10}_{-13}$ & $4000^{+2000}_{-2000}$ & $40^{+11}_{-7}$ & $31^{+9}_{-8}$ & $0.6^{+0.3}_{-0.3}$ & 2000 & 1 & 7($3 \times 10^{4}$) & 0-0(4-0) \\ 
\textbf{GW200129\textunderscore065458} & 26.5 & $1 \times 10^{5}$ & $44^{+10}_{-6}$ & $31^{+6}_{-9}$ & $1000^{+200}_{-300}$ & $37^{+9}_{-5}$ & $26^{+5}_{-7}$ & $0.19^{+0.04}_{-0.06}$ & 80 & $5 \times 10^{-4}$ & 400($2 \times 10^{4}$) & 5-0(70-27)\\ 
\textbf{GW200202\textunderscore154313} & 11.3 & $1 \times 10^{5}$ & $11^{+4}_{-2}$ & $8^{+1}_{-2}$ & $400^{+100}_{-200}$ & $10^{+4}_{-2}$ & $7^{+1}_{-2}$ & $0.09^{+0.03}_{-0.03}$ & 200 & $3 \times 10^{-4}$ & 2000($1 \times 10^{4}$) & 57-8(69-17)\\ 
GW200208\textunderscore130117 & 10.8 & 3000 & $53^{+12}_{-8}$ & $39^{+9}_{-11}$ & $2400^{+1000}_{-900}$ & $37^{+9}_{-6}$ & $27^{+6}_{-7}$ & $0.4^{+0.1}_{-0.1}$ & 40 & 0.004 & $0$(2000) & 0-0(11-0) \\ 
GW200209\textunderscore085452 & 10.0 & 20 & $56^{+14}_{-10}$ & $40^{+10}_{-10}$ & $4000^{+2000}_{-2000}$ & $34^{+9}_{-6}$ & $26^{+7}_{-7}$ & $0.6^{+0.3}_{-0.3}$ & 900 & 0.4 & 4(7000) & 0-0(3-0) \\ 
GW200219\textunderscore094415 & 10.8 & 1000 & $59^{+13}_{-9}$ & $44^{+9}_{-13}$ & $4000^{+2000}_{-2000}$ & $36^{+10}_{-6}$ & $27^{+7}_{-7}$ & $0.6^{+0.2}_{-0.2}$ & 700 & 0.2 & 7(3000) & 0-0(2-0) \\ 
\textbf{GW200224\textunderscore222234} & 19.2 & $1 \times 10^{5}$ & $53^{+9}_{-6}$ & $42^{+6}_{-10}$ & $1800^{+500}_{-700}$ & $40^{+7}_{-5}$ & $32^{+5}_{-7}$ & $0.33^{+0.07}_{-0.11}$ & 50 & 0.002 & 30($1 \times 10^{4}$) & 0-0(32-1)\\ 
\textbf{GW200225\textunderscore060421} & 13.1 & $9 \times 10^{4}$ & $24^{+5}_{-3}$ & $17^{+3}_{-5}$ & $1200^{+500}_{-500}$ & $19^{+5}_{-3}$ & $14^{+3}_{-3}$ & $0.23^{+0.08}_{-0.09}$ & 500 & 0.01 & 2000($2 \times 10^{5}$) & 3-0(68-9)\\ 
GW200302\textunderscore015811 & 10.6 & 9 & $49^{+9}_{-10}$ & $25^{+12}_{-7}$ & $1600^{+1000}_{-700}$ & $38^{+8}_{-9}$ & $19^{+8}_{-5}$ & $0.3^{+0.2}_{-0.1}$ & 7000 & 0.8 & 7000($1 \times 10^{6}$) & 0-0(50-0) \\ 
\textbf{GW200311\textunderscore115853} & 17.6 & $1 \times 10^{5}$ & $42^{+9}_{-5}$ & $33^{+5}_{-8}$ & $1200^{+300}_{-400}$ & $34^{+7}_{-4}$ & $27^{+4}_{-6}$ & $0.23^{+0.05}_{-0.07}$ & 40 & $3 \times 10^{-4}$ & 90($1 \times 10^{4}$) & 0-0(56-16)\\ 
GW200316\textunderscore215756 & 10.1 & $1 \times 10^{5}$ & $17^{+14}_{-4}$ & $9^{+3}_{-3}$ & $1100^{+400}_{-400}$ & $14^{+11}_{-4}$ & $7^{+2}_{-3}$ & $0.22^{+0.07}_{-0.08}$ & 200 & 0.005 & 500($2 \times 10^{4}$) & 2-0(59-8) \\ 

\label{table_events}
\enddata
\end{deluxetable*}
\end{longrotatetable}

\subsection{Description of the \gladeplus galaxy catalog}

In one of the analyses that we perform, the redshift information is taken from galaxy surveys for all of the events apart from GW170817, for which we assume the redshift information from its EM counterpart. For the analysis taking into account galaxy surveys we use the \gladeplus \citep{Dalya:2018cnd, Dalya:2021ewn} all-sky galaxy 
catalog that is a revised version of the first \texttt{\glade} catalog \citep{Dalya:2018cnd} containing about $22$ million  galaxies. \gladeplus  incorporates six different galaxy catalogs and surveys, namely the Gravitational Wave Galaxy Catalogue (GWGC, \citealt{GWGC}), HyperLEDA \citep{HyperLEDA}, the 2 Micron All-Sky Survey Extended Source Catalog (2MASS XSC, \citealt{2MASS_XSC}), the 2MASS Photometric Redshift Catalog (2MPZ, \citealt{2MPZ}), the WISExSCOS Photometric Redshift Catalogue (WISExSCOSPZ, \citealt{WISExSCOS}), and the Sloan Digital Sky Survey quasar catalogue from the 16th data release (SDSS-DR16Q, \citealt{SDSS_DR16_Q}) and covers the full sky with a completeness of about $20\%$ up to $800$ Mpc.\footnote{Links to the different constituent galaxy catalogs and surveys in \gladeplus are as follows: GWGC- \url{http://vizier.u-strasbg.fr/viz-bin/VizieR?-source=GWGC}, HyperLEDA-\url{http://leda.univ-lyon1.fr/}, 2MASS XSC-\url{https://old.ipac.caltech.edu/2mass/}, 2MPZ-\url{http://ssa.roe.ac.uk/TWOMPZ.html}, WISExSCOS-\url{http://ssa.roe.ac.uk/WISExSCOS.html}, SDSS-DR16Q-\url{https://www.sdss.org/dr16/algorithms/qso_catalog/}.} Most of the galaxies in the \gladeplus catalog have a redshift measurement obtained photometrically using an artificial neural network algorithm \citep{2004PASP..116..345C} with a relative error $\sigma_{z_{\rm ph}} \sim 0.033 (1+z_{\rm ph})$ \citep{WISExSCOS}. 
The peculiar velocity corrections are implemented for galaxies up to redshift $z<0.05$ using a Bayesian technique \citep{Mukherjee:2019qmm} that can capture both linear and non-linear components of the velocity field.

For our main results, we use all galaxies with measured $K_s$--band (denoted as K--band henceforth) luminosity reported in the Vega system and we assign a probability for each galaxy to be the host of a GW event that is proportional to this luminosity (\textit{luminosity weighting}). We also explore possible systematics in our results by not using  luminosity weighting and by using $B_J$--band observations; see Sec.~\ref{sec:gwcosmosys} for more details. We choose these two bands since we have found that there is a good match between the galaxy luminosity functions and the galaxy number density of the \gladeplus catalog, in particular for the K--band, see App.~\ref{app:catalogsystematics} for more details. The K--band galaxies in the \gladeplus catalog are the same as the one in \glade catalog. The galaxies in the $B_J$--band are present only in the \gladeplus catalog. 

\begin{figure*}
    \centering
    \includegraphics[scale=0.70]{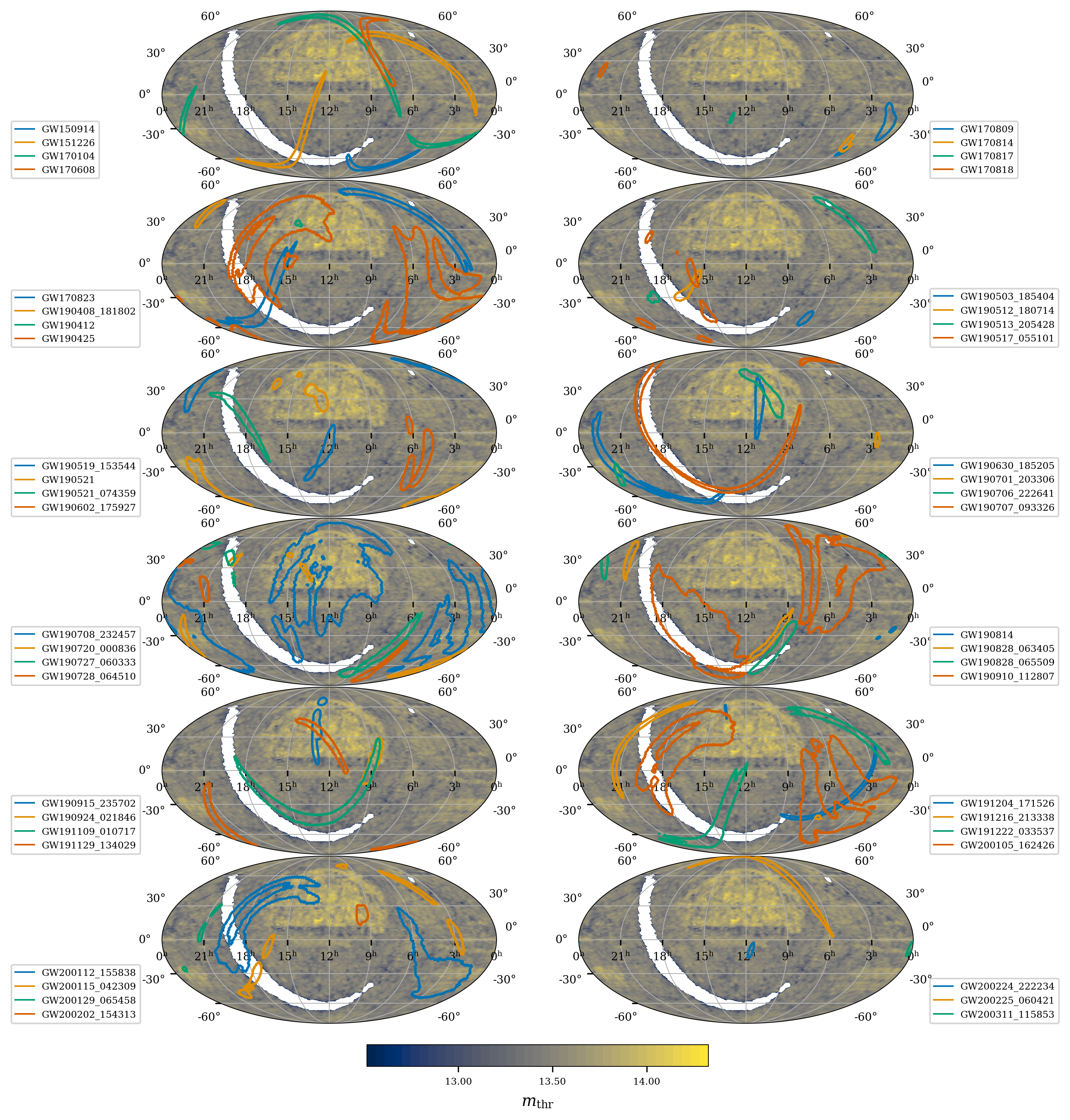}
    \caption{Skymaps showing the \gladeplus K--band apparent magnitude threshold, $m_{\rm thr}$, generated by dividing the sky into $3.35$ deg$^2$ pixels, this is the resolution used for all the events but GW190814. A mask is applied that removes from the figures all pixels with $m_{\rm thr}<12.5$ in order to improve the figure readability. Also shown are the 90\% CL sky localizations for the GW events considered in this paper.}
    \label{fig:glade+_mthr_skymap}
\end{figure*}

Fig.~\ref{fig:glade+_mthr_skymap} presents a series of skymaps showing the directional dependence of the K--band apparent magnitude threshold for the \gladeplus galaxies, in superposition with the sky localizations of the GW events included in our analysis. Outside of the galactic plane, $m_{\rm thr}\sim 13.5$ on average for the K--band while within the galactic plane region the apparent magnitude threshold is significantly lower (i.e. brighter).

We assume that the K--band absolute magnitude distribution for \gladeplus galaxies is well described by a Schechter function with parameters (reported  for $H_0=100 \hu$) $M_{\rm *,K}=-23.39$ and $\alpha{\rm _K}=-1.09$ \citep{2001ApJ...560..566K}, while for the $B_J$--band we use $M_{\rm *,B_J}=-19.66$ and $\alpha_{\rm B_J}=-1.21$ \citep{2002MNRAS.336..907N}. We set a bright cut-off high enough to include all the bright galaxies supported by the Schechter function: $M_{\rm min,K}=-27.00$ and $M_{\rm min,B_J}=-22.00$. Further, we consider all the galaxies no fainter than $M_{\rm max,K}=-19.0,M_{\rm max, B_J}=-16.5$.
These choices  correspond to all galaxies with luminosity $L>0.017L_{\rm *,K}$ and $L>0.054L_{\rm *,B_J}$, where $L_{\rm *}$ is the characteristic galaxy luminosity of the Schechter luminosity function. To calculate the rest-frame absolute magnitudes of the galaxies, for a given cosmology, we apply color and evolution corrections as reported in \citet{2001ApJ...560..566K} for the K--band and \citet{2002MNRAS.336..907N} for the $B_J$--band.

For all events, apart from GW190814, we carry out the analysis using a pixel size of 3.35 deg$^2$, while for GW190814 we use a pixel size of 0.2 deg$^2$ since the sky localization for this event was 10 times smaller than most of the others.
These values have been chosen taking into account the average number of galaxies per square degree reported in \gladeplus. Considering the bright and faint limits of the Schechter function assumed with a median apparent magnitude threshold $m_{\rm thr,K}=13.5$ for the K--band and $m_{\rm thr,B_J}=19.7$ for the $B_J$--band, we find that there are $\sim 25$ galaxies per square degree in \gladeplus reported in the K--band and $500$ galaxies per square degree reported in the $B_J$ band considered in the analysis. Note that the actual galaxy density per square degree is higher outside the galactic plane and in the region of GW190814.


For any given redshift, the \textit{completeness fraction}, which is the probability that the galaxy catalog contains the host galaxy of the GW event, $P(G|z,H_0)$, is defined as the fraction of galaxies with absolute magnitudes brighter than the absolute magnitude threshold (calculated from $m_{\rm thr}$), namely
\begin{equation}
    P(G|z,H_0) = \frac{\int^{L_{\rm max}}_{L_{\rm thr}(m_{\rm thr},z,H_0)} \phi(L) L {\rm d}L}{\int^{L_{\rm max}}_{L_{\rm min}} \phi(L) L {\rm d}L}.
    \label{eq:cat_comp}
\end{equation}
Here $\phi(L)$ is the assumed galaxy luminosity function, $L_{\rm min}$ and $L_{\rm max}$ are the minimum and maximum luminosity corresponding to $M_{\rm max}$ and $M_{\rm min}$ and $L_{\rm thr}$ is the threshold luminosity for detection, calculated from $m_{\rm thr}$.

\begin{figure}
    \centering
    \includegraphics[scale=0.9]{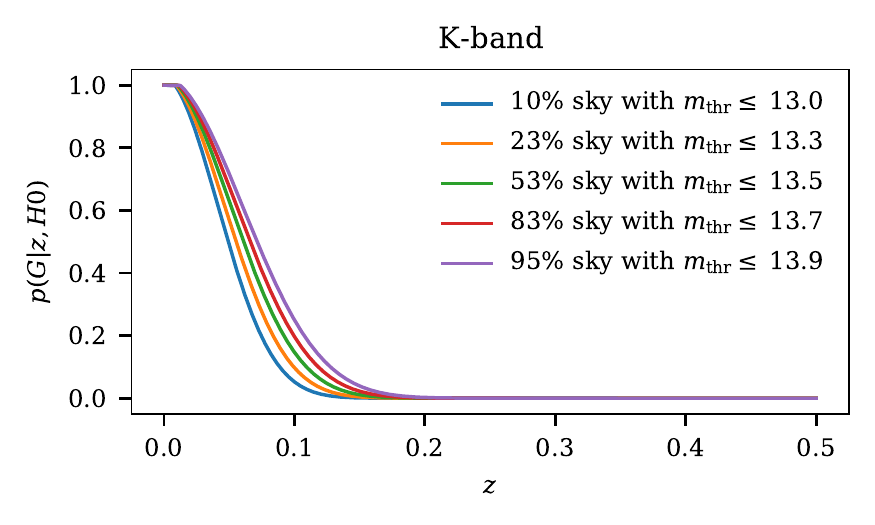}
    \includegraphics[scale=0.9]{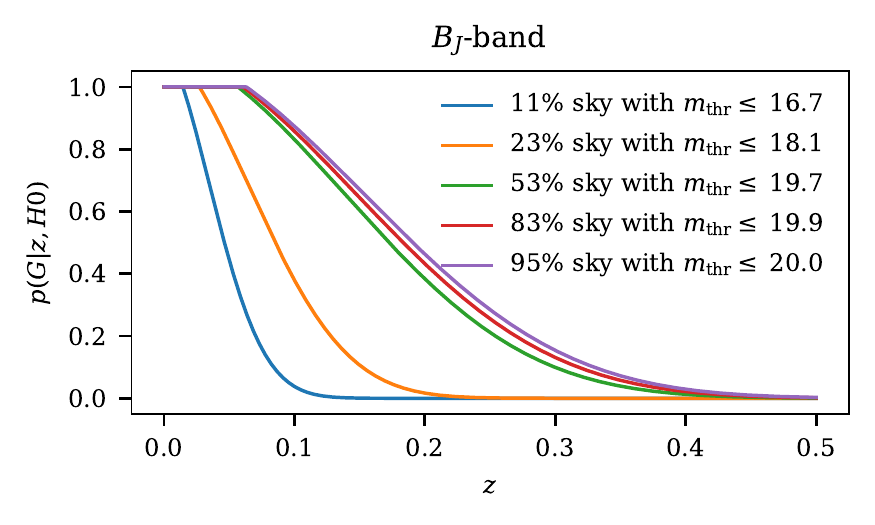}
    \caption{\textit{Top}: Completeness fraction of \gladeplus in the K--band, indicating the probability that the catalog contains the host galaxy of a GW event, as a function of redshift for $H_0=67.9 \hu$ and $\Omega_{\rm m}=0.3065$. The different lines are calculated for a given K--band apparent magnitude $m_{\rm thr}$ threshold indicated in the legend. The legend also indicates the fraction of the sky, computed by dividing the sky in equal sized pixels of $3.35$ deg$^2$, for which the apparent magnitude threshold is brighter than the one reported in the legend. The fraction of pixels with no galaxies is $\sim 5\%$ for the $B_j$ and K bands. \textit{Bottom}: Same as the top panel but for the $B_J$--band.}.
    \label{fig:glade+_completeness}
\end{figure}

Fig.~\ref{fig:glade+_completeness} shows the completeness fraction of the \gladeplus catalog, in the K--band and $B_J$--band respectively, as a function of redshift and for different values of $m_{\rm thr}$, assuming a fiducial cosmology with $H_0=67.9\, \hu$ and $\Omega_{\rm m}=0.3065$ \citep{Planck:2015fie}. As can be seen from the figure, the \gladeplus catalog is less complete in the K--band than in the $B_J$--band, but we decide to use the K--band data for our main results as they are better described by the Schechter function assumed in our analysis; see App.~\ref{app:catalogsystematics}. 

For \gladeplus systematic uncertainties of the photometric redshift reconstruction are inside the statistical errors associated to each galaxy \citep{2MPZ} with a very small percentage of outliers.
We do not consider deeper galaxy surveys with a restricted sky area footprint, such as the \texttt{DES Y1} survey \citep{2018ApJS..235...33D} or \texttt{DESI Legacy Imaging Survey} \citep{2019ApJS..242....8Z} that are supposed to be complete up to redshift $z \sim 1$, since we decide to employ the same all-sky galaxy catalog for all of our events. Moreover, color corrections and photometric redshift reconstruction might need particular attention with deeper galaxy surveys.

\section{Results \label{sec:results}}

The first analysis that we present in Sec.~\ref{sec:cosmopop} will focus on the impact of the BBH population source masses on inference of the cosmological parameters, using the formalism discussed in Sec. \ref{sec:icarogw}. 
This analysis uses no galaxy catalog information; instead constraints on the cosmological parameters will be inferred from the mass scale set by the source mass distribution. We use only the BBHs detections as they are the majority of the sources entering in our cosmological analysis and because a joint description of NSBH, BNS and BBHs is uncertain.

The second analysis in Sec.~\ref{sec:gwcosmo} fixes the source mass distribution and uses redshift information derived from galaxy catalogs, based on the formalism discussed in Sec. \ref{sec:gwcosmo_method}. Unless stated otherwise, all the figures are generated with a uniform prior on $H_0$, credible intervals are reported as maximum posterior and 68.3\% highest density intervals. We use a flat-in-log prior on $H_0$ only when quoting results combined with GW170817 and its EM counterpart. 
\subsection{Implications of population assumptions for cosmology \label{sec:cosmopop}}

We jointly estimate population-related GW parameters and cosmological parameters using BBH events, since these are the majority of the GW events observed to date.
We use the \Nevicarogw  BBH detected events with SNR $>11$. We exclude from this analysis GW190814 \citep{Abbott:2020gyp} given the current uncertainty on the nature of the secondary object in this system.

We consider two cosmological models:  \textit{(i)} a flat $w_0$CDM model with  \textit{wide priors\/} on the Hubble constant $H_0$, matter density $\Omega_{\rm m}$ and dark energy equation of state (EoS) $w_0$ parameter, and \textit{(ii)} a flat $\Lambda$CDM Universe with a fixed value of $\Omega_{\rm m}=0.3065$ \citep{Planck:2015fie} and dark energy EoS parameter $w_0=-1$ and with a \textit{restricted prior} in the $H_0$ tension region ($H_0 \in [65,77]\, \hu$). We refer to model \textit{(i)} as the $w_0$CDM model, and to model \textit{(ii)} as the $H_0$-tension model. We also adopt wide priors on the hyper-parameters of the GW source mass distribution and its merger rate evolution as described in App.~\ref{app:priors}. For all the phenomenological mass models assumed we obtain posteriors on the source mass distribution and merger rate parameters which are compatible with previous population studies \citep{Abbott:2020gyp, O3bPops_PLACEHOLDER} and the latest studies with O3b data \citep{O3bCat_PLACEHOLDER}. See App.~\ref{app:fullicarogw} for more details.

As evident from values of Bayes factor reported in Table~\ref{tab:BF_mass_cosmo}, we do not find any preference of the data in supporting any one of the cosmological models ($w_0$CDM model or the $H_0$-tension model) considered in the analysis. As we will see later, this is because the posteriors on $\Omega_{\rm m}$ and $w_0$ are not constrained by the GW observations and the error on the $H_0$ estimation extends beyond the $H_0$ tension region.

\begin{table}[t]
  \centering
  \begin{tabular}{p{5.0cm}  c r }
    \hline
    {\bf Mass model} & \textbf{$\log_{10}{\cal B}$}  \\ 
    \hline\hline
    \PL & $\PLPLlogtenBfactCOSMO$\\
    \PLG & $\PLGPLGlogtenBfactCOSMO$\\
    \BPL & $\BPLBPLlogtenBfactCOSMO$ \\
    \hline
  \end{tabular}
  \caption{Logarithm of the Bayes factor comparing runs that adopt the same source mass model but different cosmologies: wide priors (for a general $w_0$CDM  cosmology) versus restricted priors (in the $H_0$ tension region).}
  \label{tab:BF_mass_cosmo}
\end{table}

In Table~\ref{tab:BF_mass} we report the Bayes factors computed between different mass models, for the case of wide priors on the $w_0$CDM cosmological parameters.
Consistent with \citet{Abbott:2020gyp, O3bPops_PLACEHOLDER}, we find that, even if we allow the cosmological parameters to vary with wide priors, the \PL model is still strongly disfavored with respect to the \PLG and \BPL models, by a factor $\sim 100$. This result is consistent with the fact that, as indicated in Fig.~\ref{fig:representation_tab1}, the source mass distribution contains more structure than a simple \PL model. As motivated in \citet{Abbott:2020gyp}, this comparatively poor fit for the \PL model is due to the inability of this model to capture a moderate fraction of detected events with high masses, while predicting a large fraction of detected events with lower masses. Using the reduced set of signals with SNR $>11$, we do not find any compelling evidence to prefer the \PLG model over the \BPL model. 

\begin{table}[h]
  \centering
  \begin{tabular}{p{5.0cm}  c r }
    \hline
    {\bf Mass model} & \textbf{$\log_{10}{\cal B}$}  \\ 
    \hline\hline
    \PL & $\PLGPLlogtenfact$\\
    \PLG & $\PLGPLGlogtenfact$\\
    \BPL & $\PLGBPLlogtenfact$ \\
    \hline
  \end{tabular}
  \caption{Logarithm of the Bayes factor between the different mass models and the \PLG model preferred by the data, for the case of a $w_0$CDM cosmology with wide priors.}
  \label{tab:BF_mass}
\end{table}

\begin{figure}
    \centering
    \includegraphics[scale=1]{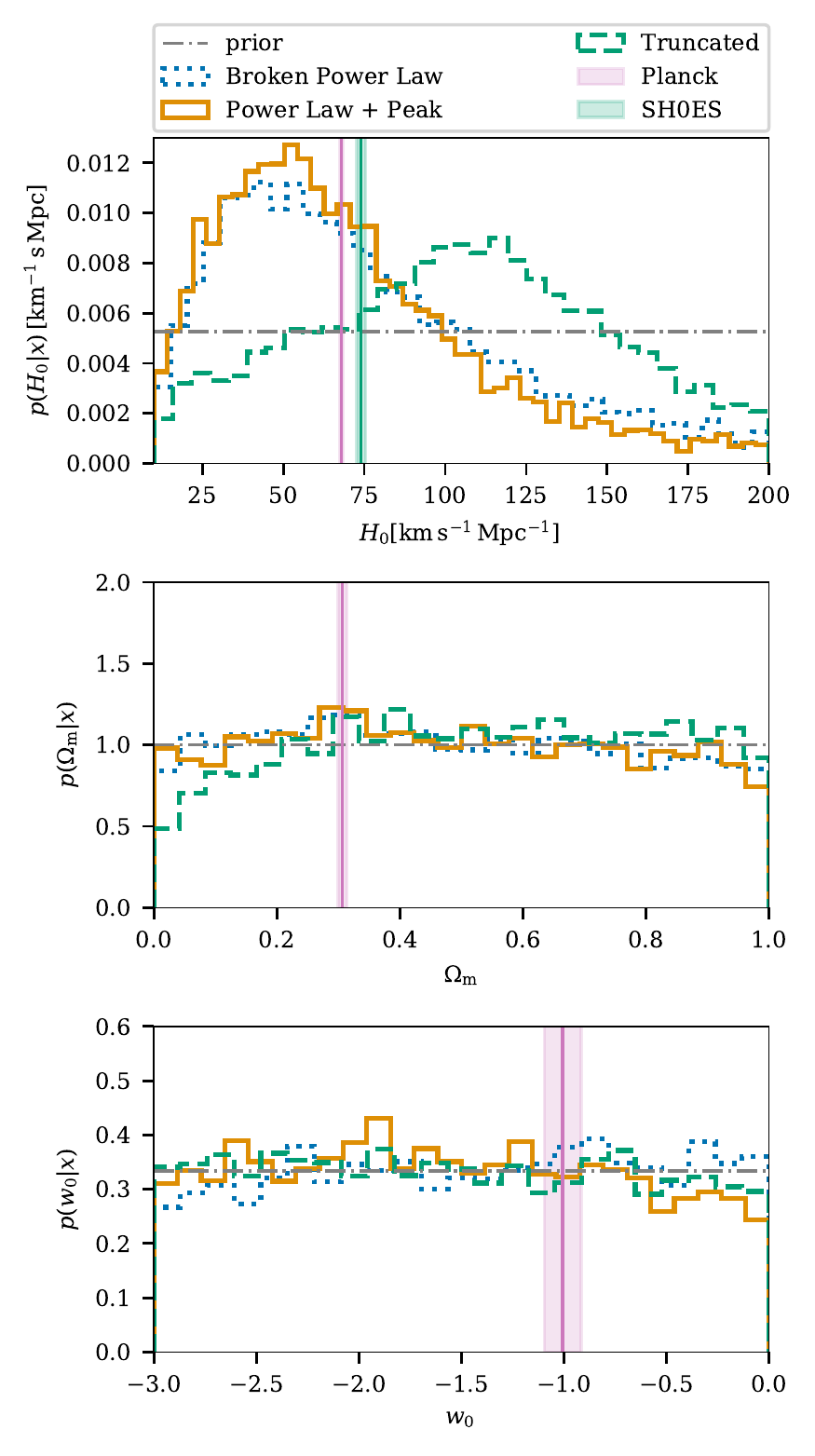}
    \caption{\textit{Top panel}: Marginal posterior distribution for $H_0$. \textit{Middle panel}: Marginal posterior distribution for $\Omega_{\rm m}$. \textit{Bottom panel}: Marginal posterior distribution for $w_0$. In each panel the different lines indicate the 3 phenomenological mass models. The solid orange line identifies the preferred \PLG model. The pink shaded areas identify the 68\% CI of the cosmological parameters inferred from measurements from the CMB \citep{Planck:2015fie} (apart for $w_0$ that is reported at 95\% CI) and the green shaded area in the top panel shows the value of the Hubble constant measured in the local Universe \citep{Riess:2019cxk}.}
    \label{fig:recap_w_190521}
\end{figure}

The marginal posterior distributions that we obtain for the cosmological parameters  $H_0$, $\Omega_{\rm m}$ and $w_0$ are shown in Fig.~\ref{fig:recap_w_190521} for each phenomenological mass model.
As anticipated by our Bayes factor results, we find that with the current BBH GW events we cannot constrain the values of these three cosmological parameters, as we obtain broad and uninformative posteriors.

\begin{figure*}
    \centering
    \includegraphics[scale=1.0]{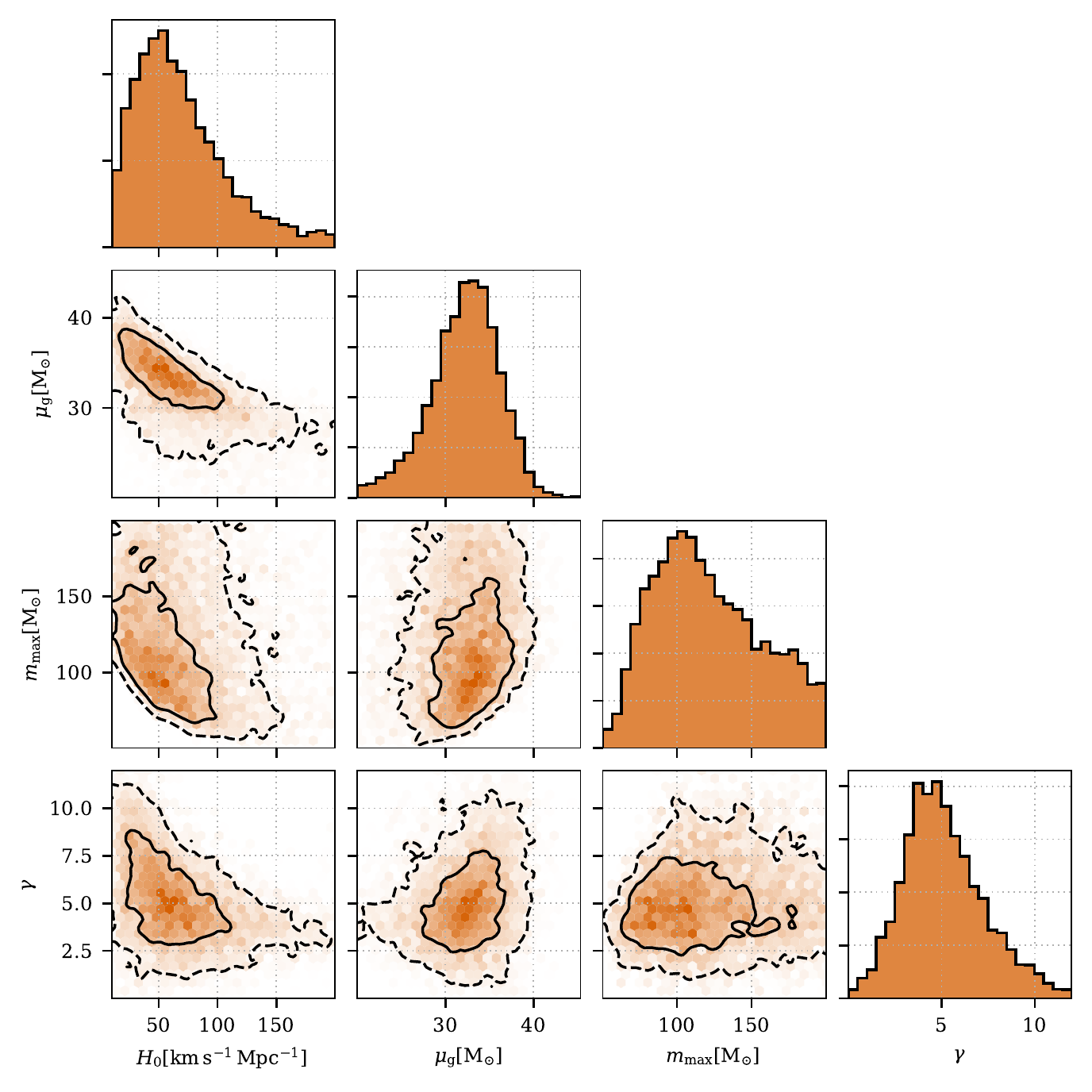}
    \caption{Posterior probability density for $H_0$ and the population parameters $\mu_g, m_{\rm max}$ and $\gamma$, governing the position of the Gaussian peak, the upper end of the mass distribution and the merger rate evolution in the \PLG mass model. The solid and dashed black lines indicate the 50\% and 90\% CL contours.}
    \label{fig:PLG_w0flatLCDM_corners}
\end{figure*}

With the \PLG we estimate $H_0=\HnoticarogwPLG \hu$, while for the \BPL model we estimate $H_0=\HnoticarogwBPL \hu$.
These constraints on $H_0$, as we will see later, arise from the ability of these models to fit an excess of BBHs with masses around $35\, \Msol$ which sets a scale for the redshift distribution of BBHs.

We discuss this effect further using the \PLG model. Fig.~\ref{fig:PLG_w0flatLCDM_corners} shows the joint posterior distribution between the cosmological parameters and the parameters $\mu_{\rm g}$ and $m_{\rm max}$ defined in Eq.~(\ref{eq:PLG}), which govern the position of the BBH Gaussian excess and the upper end of the source primary mass distribution respectively.

The presence of a peak in the BBH source mass distribution allows us to set a characteristic source mass scale, which informs $H(z)$ and allows us to exclude higher values of $H_0$.
Marginalizing over the cosmological parameters, we obtain a central value of $\mu_{\rm g}=\mugcarogw\, \Msol$ for the peak position of the Gaussian BBH excess. 
On the other hand, the disfavoured \PL model shows support at higher $H_0$. This result is due to the fact that the \PL model is not able to adequately fit the presence of massive binaries while producing an excess of BBHs with masses $\sim 40\, \Msol$ in the detector frame. For this reason, higher $H_0$ values are more supported since those values place events at higher redshifts, thus reducing their source masses. 

\begin{figure}
    \centering
    \includegraphics[scale=1]{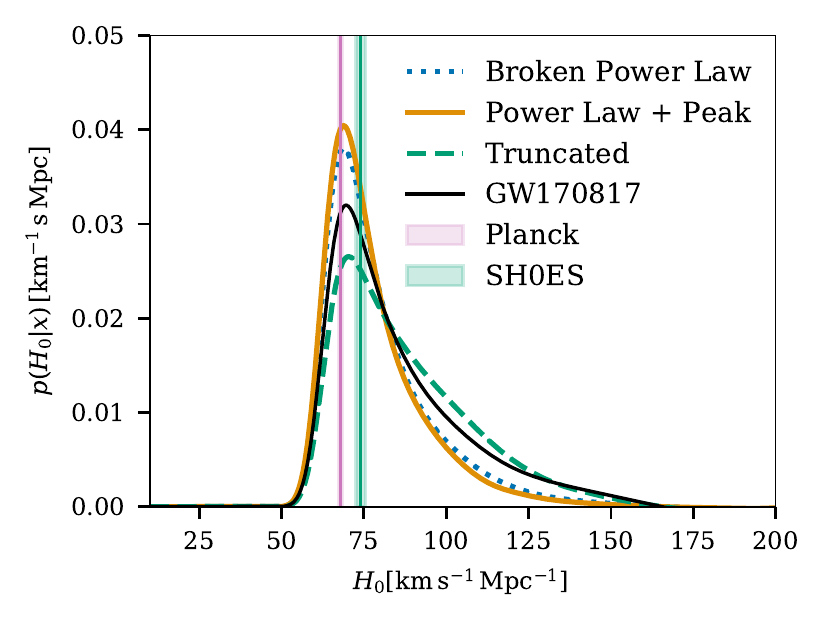}
    \caption{Posterior distributions for $H_0$ obtained by combining the $H_0$ posteriors from the \Nevicarogw\, BBH detections and the $H_0$ posterior inferred from the bright standard siren GW170817. The pink and green shaded areas identify the 68\% CI constraints on $H_0$ inferred from the CMB anisotropies \citep{Planck:2015fie} and in the local Universe from SH0ES \citep{Riess:2019cxk} respectively.}
    \label{fig:PLG_with_GW170817}
\end{figure}

When we combine the $H_0$ posteriors from the three mass models with the $H_0$ inferred from the bright standard siren GW170817 (see Fig.~\ref{fig:PLG_with_GW170817}), we find a value of $H_0=\HnotnoticarogwwithBNSlogPLG\, \hu$ for the \PLG model and $H_0=\HnotnoticarogwwithBNSlogBPL\, \hu$ for the \BPL model. These results represent an improvement of \HnoticarogwimproGWTConePLG\% and \HnoticarogwimproGWTConeBPL\%  respectively compared with the  $H_0$ value reported in \citet{Abbott:2019yzh} that made use of GW170817 and six BBH detections from O2, with redshift information inferred from galaxy catalogs. 
For the \PL model, we obtain $H_0=\HnotnoticarogwwithBNSlogPL \hu$. These results are obtained assuming a redshift independent mass distribution. Considering a redshift dependence of the mass distribution, can degrade the constraints. 

\subsection{Results using galaxy catalog information \label{sec:gwcosmo}}

We now discuss constraints on $H_0$ when we fix the source population model but employ galaxy surveys to infer statistical redshift information using the pixelated \texttt{gwcosmo} code \citep{Gray:2019ksv}. Our analysis incorporates \Nevgwcosmo GW events, comprising \Nevicarogw BBH detections, GW190814, the two BNS events GW170817 and GW190425, and the two NSBH events GW200105 and GW200115. We include all galaxies of the \gladeplus catalog that lie inside the 99.9\%  estimated sky area of each event. We use the \gladeplus K--band data in this analysis, adopting luminosity weights for each galaxy. For a more in-depth discussion about the impact of our BH population assumptions and choice of photometric bands, see Sec.~\ref{sec:gwcosmosys}. 

To describe the distribution of BH primary masses, we use a \PLG source mass model where we fix population parameters to the median values obtained in the joint cosmological and population analysis described in Sec.~\ref{sec:cosmopop}. For the rate evolution we adopt $\gamma=4.59$, $k=2.86$ and $z_p=2.47$, while for the \PLG model we use $\alpha=3.78$, $\beta=0.81$, $m_{\rm max}=112.5 \, \Msol$, $m_{\rm min}= 4.98 \, \Msol$, $\delta_m=4.8 \, \Msol$ $\mu_{\rm g}=32.27 \, \Msol$, $\sigma_{\rm g}=3.88\, \Msol$ and $\lambda_{\rm g}=0.03$. 
For the NS source mass model we consider a uniform distribution between $m_{\rm min}1 \Msol$ and $m_{\rm max}=3 \Msol$ consistently with \citep{O3bPops_PLACEHOLDER}.
We evaluate GW selection effects using LIGO and Virgo sensitivities during the O1, O2, and O3 runs. 

\begin{figure*}
\centering
    \hspace*{-2.0cm}
    \includegraphics[scale=1.15]{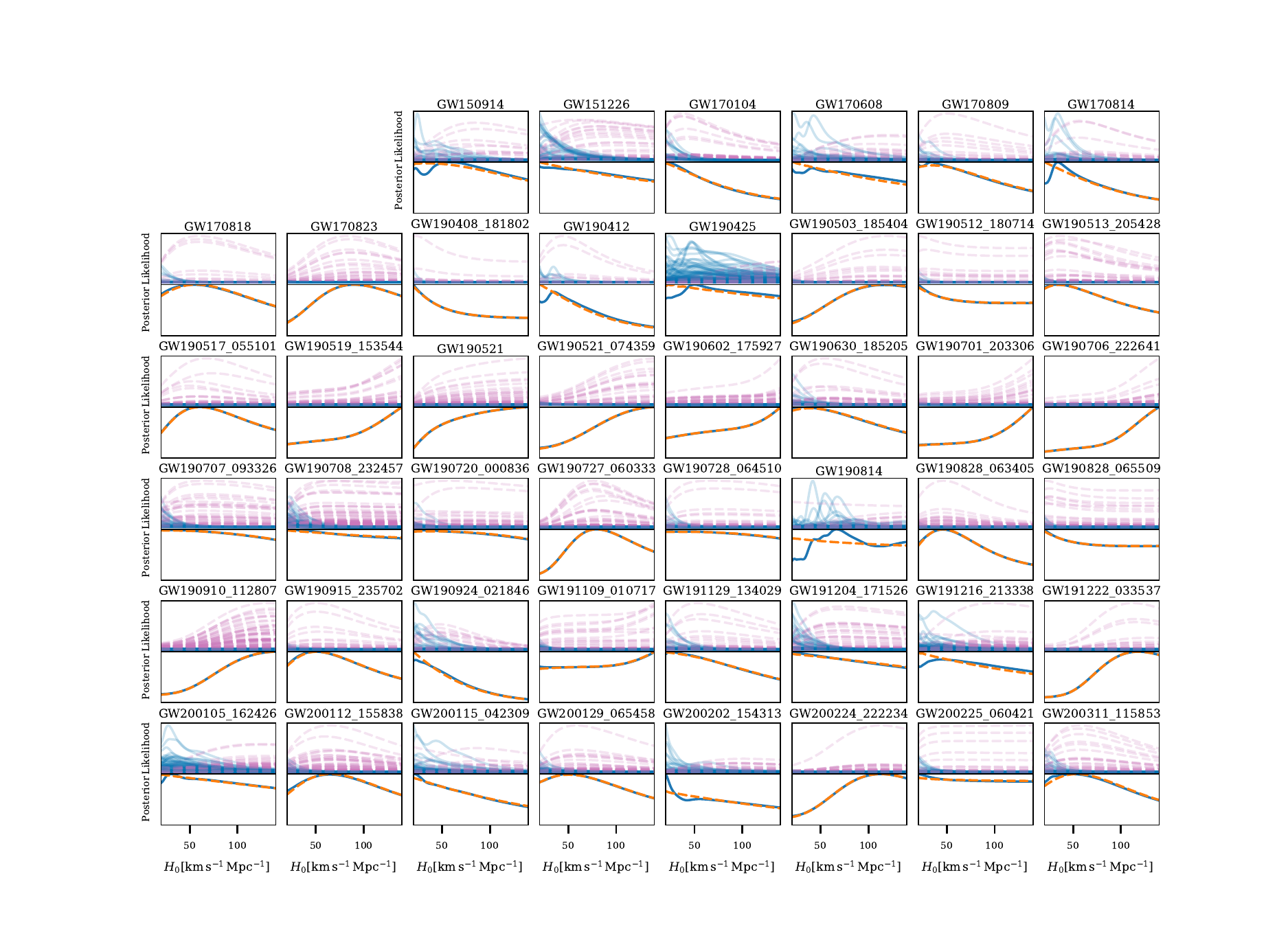}
    \caption{Plots reporting the results on the $H_0$ inference for each event using the \gladeplus K band and luminosity weighting. Two panels are shown for each event. \textit{Top panels:} Hierarchical likelihood under the hypothesis, $G$, that the host galaxy is in the catalog (blue solid lines) and under the hypothesis, $\bar{G}$, that the host galaxy is \textit{not} in the catalog (pink dashed lines). The different lines shown in each panel correspond to the different pixels within the sky localization area for each event. \textit{Bottom panels:} The blue solid line shows the posterior obtained by summing the terms corresponding to the in-catalog and out-of-catalog hypotheses. The orange dashed line shows the posterior obtained by assuming a galaxy catalog with null completeness. In this case the $H_0$ inference comes entirely from the population assumptions. This plot is intended to show which event is informative on the $H_0$ value and whether the information is coming from population assumptions or galaxy catalog contribution.}
    \label{fig:posteriors}
\end{figure*}

In Fig.~\ref{fig:posteriors} (page \pageref{fig:posteriors}) we show the posteriors for all of the GW events considered in this analysis for the K--band.
For many  of the O3 events, the $H_0$ inference is dominated by the likelihood based on the hypothesis that the host galaxy is \textit{not} in the catalog (referred to as out-of-catalog). The out-of-catalog term dominates for sources that are localized at redshifts at which the \gladeplus galaxy catalog has low completeness fraction (see Fig.~\ref{fig:glade+_completeness}). This is the case for most of the GW sources which are BBHs observed at large luminosity distances. Another interesting trend observed in Fig.~\ref{fig:posteriors} is that, for lower values of $H_0$, the in-catalog likelihood terms tend to dominate because for low $H_0$ values the GW events are placed at smaller redshifts where the galaxy catalog is more complete, as shown in Fig.~\ref{fig:glade+_completeness}.

For most of these events, the number of galaxies present in the sky localization volume is large enough that the redshift information is still dominated by population assumptions (Section 5.2). 
GW190814 is the only event for which there is a sufficiently small number of galaxies in its sky localization area of about $18$ deg$^2$. Its small area makes this event partially more informative on the value of $H_0$ in comparison to the other GW events. We can see in Fig.~\ref{fig:posteriors} that, out of all the GW events, the most informative posterior on $H_0$ (compared to the zero galaxy catalog completeness posterior) is from GW190814, provided that the luminosity weighting scheme is applied.
We have verified that the $H_0$ posterior with the K--band and using luminosity weights does not depend on the faint end magnitude limit used for the analysis.
For this event, we infer an $H_0$ constraint of $\HnotgwcosmoGWnof\, \hu$ (MAP and HDI). We quote the maximum a posteriori probability (MAP) and the corresponding highest density interval (HDI) values in the analysis.

\begin{figure*}
    \centering
    \includegraphics[scale=0.9]{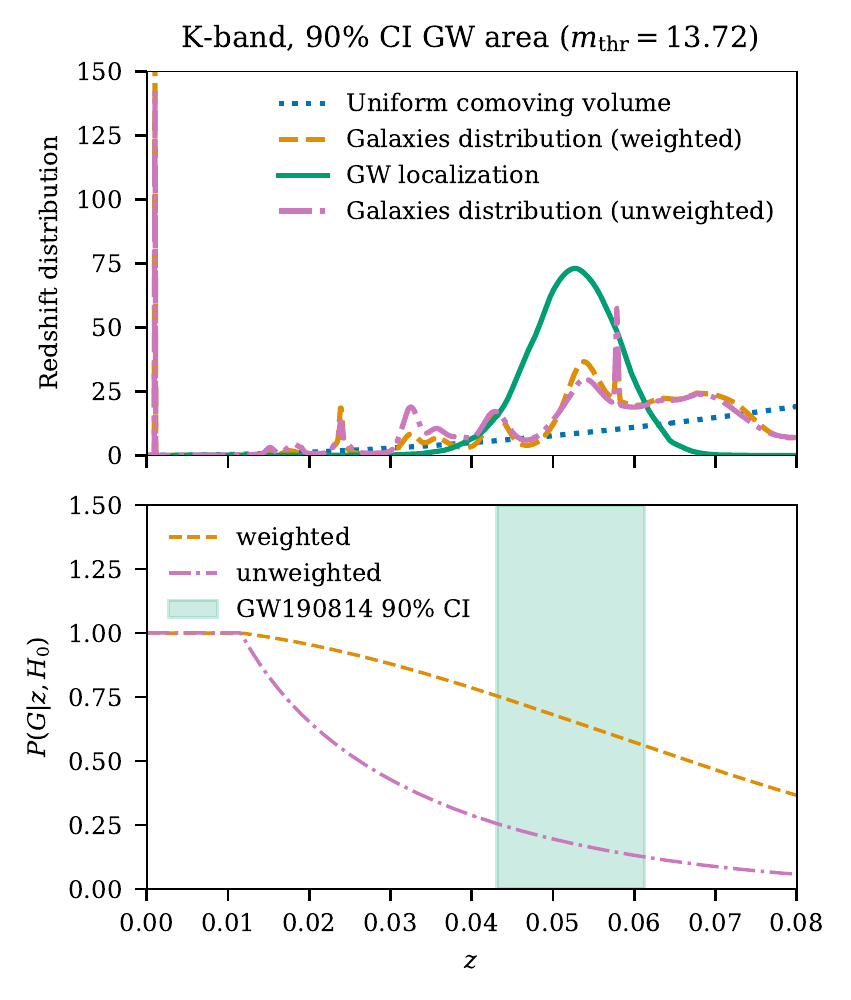}
    \includegraphics[scale=0.9]{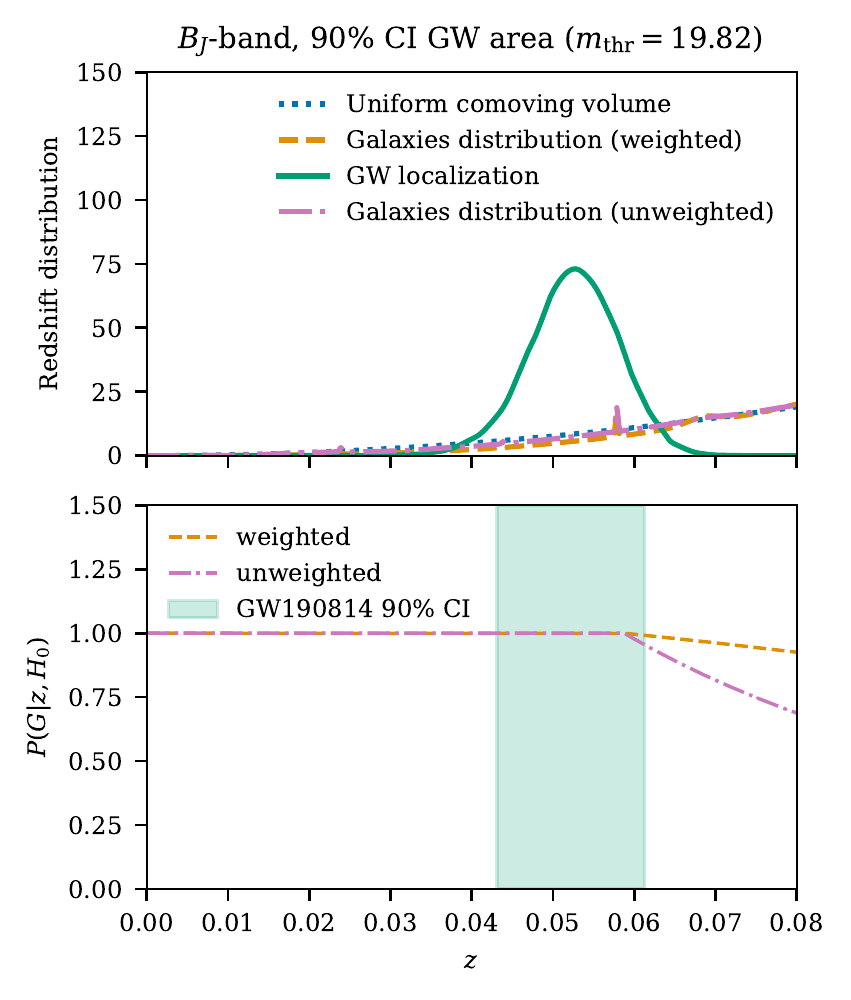}
    \caption{ \textit{Top panels:} Distribution of galaxies observed in the \gladeplus K(left panels) and $B_J$ (right panels) bands as a function of redshift $z$ with and without galaxy luminosity weights, compared with the GW190814 redshift localization in its 90\% CI sky area, assuming a cosmology with $H_0=67.9 \hu$ and $\Omega_{\rm m}=0.3065$ (green line) and the predicted redshift distribution for a prior that is uniform in comoving volume (blue dashed line). The redshift distribution is not intended as representative of the reconstructed galaxy distribution since it is calculated by stacking each galaxy redshift localization assumed as a normal distribution. This procedure only serves to give a rough idea where the $H_0$ contribution is coming from. \textit{Bottom panels:} Completeness calculated using the K and $B_J$ band with and without application of the luminosity weighing scheme.}
    \label{fig:GW190814A_overdensity}
\end{figure*}

Fig.~\ref{fig:GW190814A_overdensity} shows the redshift distribution of galaxies in the $90\%$ CI sky area of GW190814 (top panel) and the galaxy catalog completeness (bottom panel), compared to the predicted distribution for a prior that is uniform in comoving volume. We observe that for the K--band the $H_0$ support results from an excess of galaxies, with respect to the uniform in comoving volume prior, around $z\sim 0.051$. Switching off the luminosity weighting assumption decreases the contribution of this excess of galaxies since the completeness is estimated to be lower. The same excess is not visible in the $B_J$--band as more galaxies are reported in this band and some luminous galaxies with measured K--band apparent magnitudes do not have measured apparent magnitudes for the $B_J$--band.

Despite the cases where there is a significant in-catalog contribution, the final $H_0$ result is nevertheless dominated by the BBHs population assumptions which are contributing to the out-of-catalog likelihood terms (when the galaxy catalog is not complete) and in the in-catalog terms when a large number of galaxies is present in the GW sky localization volume.

\begin{figure}
    \centering
    \includegraphics[scale=0.9]{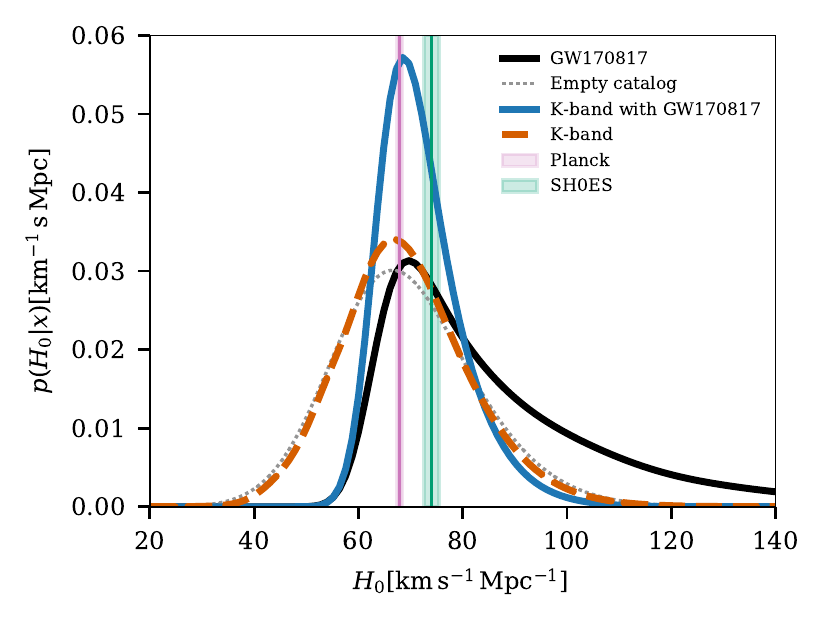}
    \caption{Hubble constant posterior for several cases. Gray dotted line: posterior obtained using all dark standard sirens without any galaxy catalog information and fixing the BBH population model. Orange dashed line: posterior using all dark standard sirens with \gladeplus K--band galaxy catalog information and fixed population assumptions. Black solid line: posterior from GW170817 and its EM counterpart. Blue solid line: posterior combining dark standard sirens and \gladeplus 
    K--band catalog information (orange dashed line) with GW170817 and its EM counterpart (black solid line). The pink and green shaded areas identify the 68\% CI constraints on $H_0$ inferred from the CMB anisotropies \citep{Planck:2015fie} and in the local Universe from SH0ES \citep{Riess:2019cxk} respectively.}
    \label{fig:gwcosmo_result_Kband}
\end{figure}

In Fig.~\ref{fig:gwcosmo_result_Kband} we show the combined $H_0$ posterior inferred from all of the GW events and for several different scenarios. By using all of the dark sirens, together with K--band galaxy information from \gladeplus, we obtain a value of $H_0 = \HnotgwcosmoKband\, \hu$. This value is strongly dominated by the BH population assumptions, as can be seen in Fig.~\ref{fig:gwcosmo_result_Kband}. The $H_0$ value obtained from population assumptions alone (Empty catalog case in Fig.~\ref{fig:gwcosmo_result_Kband}) is $H_0 = \HnotgwcosmoEmpty\, \hu$.
When we combine the galaxy catalog measurement with the result from the bright standard siren GW170817, we obtain $H_0=\HnotgwcosmoBNSKband \hu$. This value represents an improvement of \HnotgwcosmoimproGWTConeKband\% with respect to the corresponding result obtained with GWTC--1 \citep{Abbott:2019yzh}, and an improvement of \HnotgwcosmoimproBNSKband\% with respect to the result of $H_0 = 69^{+17}_{-8} \hu$ obtained using only the GW170817 event \citep{Abbott:2019yzh} .

\section{Discussion \label{sec:discussion}}

\subsection{Considerations for the BBH-based population analysis}

We have shown how population assumptions on BBH formation dominate the inference on cosmological parameters and, in particular, we have seen how the presence of an excess of BBHs with primary masses between $30\, \Msol$ and $40\, \Msol$ \citep{Farr:2019twy} sets a scale for the BBH redshifts, thus allowing for a weak constraint on $H_0$.



\begin{figure}
    \centering
    \includegraphics[scale=1.0]{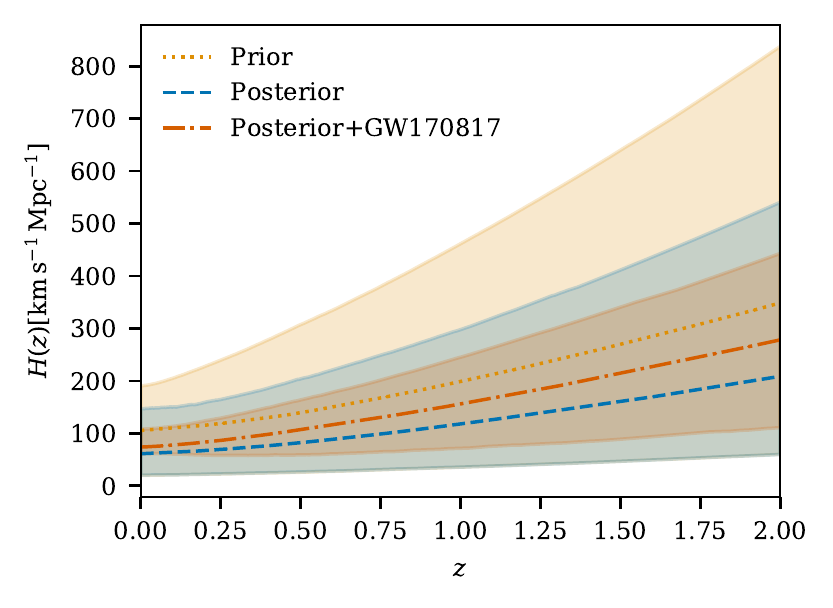}
    \caption{Evolution of the Hubble parameter predicted from the most preferred mass model \PLG (blue lines). The yellow shaded area indicates the 90\% CL contours identified by the uniform priors on $H_0$, $\Omega_{\rm m}$ and $w_0$ while the blue shaded area indicates the 90\% CL contours from the posterior of the preferred mass model. The dashed lines indicate the median of the prior and posterior for $H(z)$ respectively.}
    \label{fig:PLG_Hz}
\end{figure}

In the BBH-based population analysis without GW170817, the $(H_0, \Omega_{\rm m},w_0)$ parameters are not constrained. In Fig.~\ref{fig:PLG_Hz} we portray these constraints on the expansion rate of the Universe, $H(z)$. 
The best constraint that we obtain on the expansion rate of the Universe has a value, and uncertainty (median and symmetric 90\%\, CI), of $\HzbestBNS \hu$ at redshift $z=0$ if we include the bright siren GW170817, and $\Hzbest \hu$ at redshift $z\sim \zbest$ without it.

\subsection{Considerations on the catalog analysis}
\label{sec:gwcosmosys}


As already discussed in Sec.~\ref{sec:gwcosmo}, the $H_0$ inference is dominated by the population assumptions of the underlying BH mass distribution. In particular, as shown in  Fig.~\ref{fig:PLG_w0flatLCDM_corners}, the population parameter that is most strongly correlated with the value of $H_0$ is the position of the BHs excess $\mu_{\rm g}$. 

\begin{figure}
    \centering
    \includegraphics[scale=1.]{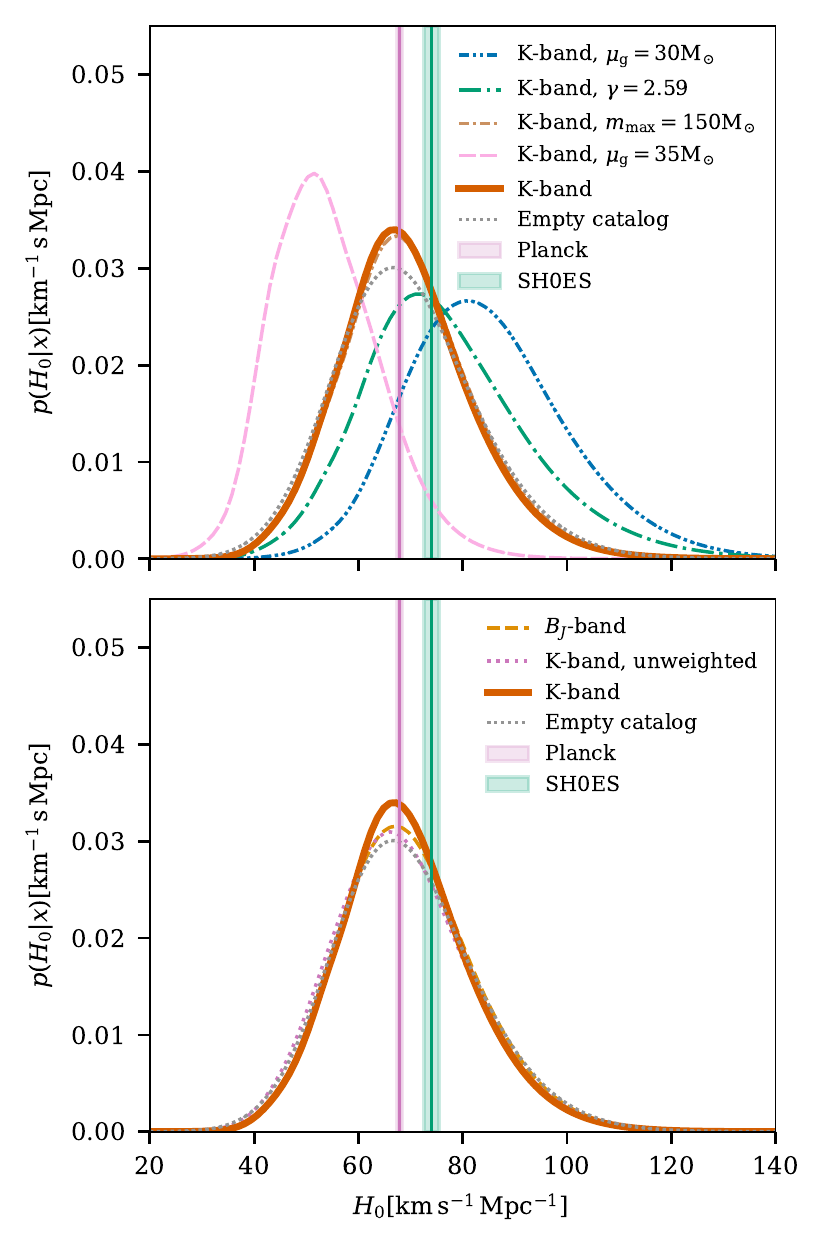}
    \caption{Systematic effects on the inference of the Hubble constant due to the choice of different values for the mean $\mu_g$ of the Gaussian component in the source mass model, and other population model parameters (upper panel) and different choices for the luminosity band and weighting scheme adopted for the \gladeplus galaxy catalog (lower panel). The pink and green shaded areas identify the 68\% CI constraints on $H_0$ inferred from the CMB anisotropies \citep{Planck:2015fie} and in the local Universe from SH0ES \citep{Riess:2019cxk} respectively.}
    \label{fig:gwcosmo_result_sys}
\end{figure}

In Fig.~\ref{fig:gwcosmo_result_sys} we show the $H_0$ posterior computed with different choices of $\mu_{\rm g}$ and fixing the remaining parameters. The values of $\mu_{\rm g}$ are spaced by $\sim 2.5 \Msol$, which is roughly the uncertainty identified in Sec.~\ref{sec:icarogw}.
It can be seen that the value of $\mu_{\rm g}$ has a strong impact on the inference of $H_0$. For values of $\mu_{\rm g}$ higher than the median value of $32.55 \Msol$, the posterior supports low $H_0$ values as we need GW events to be at a lower redshift to explain the excess of BHs at higher masses. On the other hand, for $\mu_{\rm g}<32.55\, \Msol$, the posterior supports higher $H_0$ value in order to place events at higher redshift compatible with an excess of BHs at lower masses.

We also explored the effect of raising the maximum mass of the black holes to $m_{\rm max}=150 \, \Msol $. As can be seen in Fig.~\ref{fig:gwcosmo_result_sys}, raising $m_{\rm max}$ does not have a significant effect on the $H_{0}$ posterior since only few events are present at these masses. One more parameter for which we explored the effect of its variation is the $\gamma$ parameter in the rate evolution model. In the same plot one can see the $H_{0}$ posterior for $\gamma = 2.59$. This parameter has a stronger effect on the $H_{0}$ posterior, making the posterior less informative and at the same time moving its peak to higher values. 


The galaxy catalog brings additional information only for GW190814, due to the much better sky localization ($\sim 18$ deg$^2$) for this event; this has the effect of providing more support for the $H_0$ tension region. 

\begin{figure*}
\centering
    \hspace*{-2.6cm}
    \includegraphics[scale=1.2]{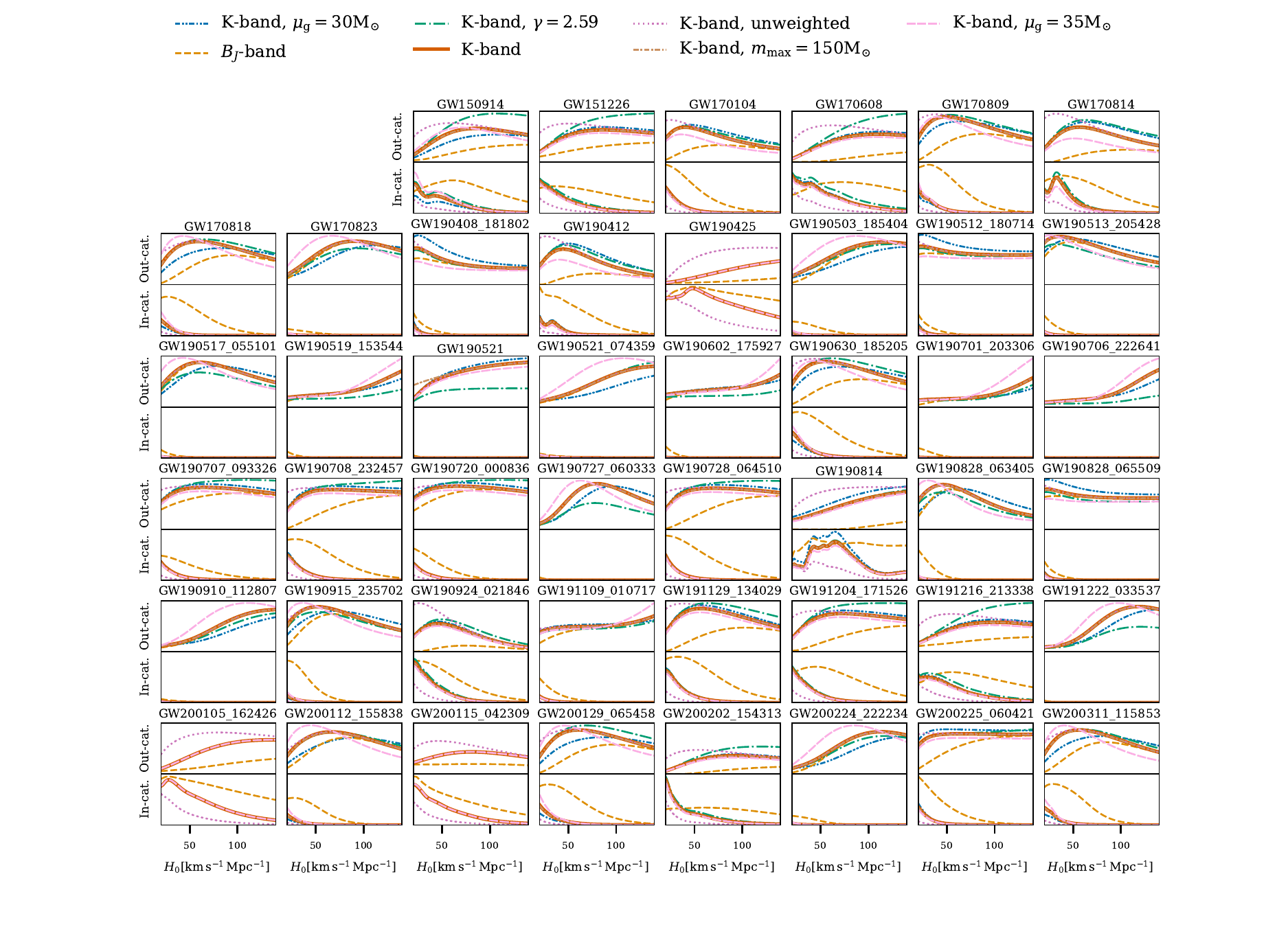}
    \caption{Plots showing, for each event, the hierarchical likelihood as a function of $H_0$ marginalized over sky localization as a function of different population assumptions and using the \gladeplus K-band. The y-axes of each panel start from zero and have the same normalization. Two panels are shown for each event. \textit{Top panels:} Hierarchical likelihood under the hypothesis that the host galaxy is \textit{not} in the catalog, for several different population assumptions. \textit{Bottom panels:} Hierarchical likelihood under the hypothesis that the host galaxy  \textit{is} in the catalog, again for several different population assumptions. This plots how population and galaxy catalog treatment affect the information on the $H_0$ value for each event.}
    \label{fig:posteriors_pop}
\end{figure*}

In Fig.~\ref{fig:posteriors_pop}, we show how population assumptions impact the hierarchical likelihood calculation as a function of $H_0$, for the hypotheses that the host galaxy is (or \textit{is not}) \textit{inside} the catalog. 
Population assumptions strongly impact the out-of-catalog term of the likelihood, which is the dominant contribution to the $H_0$ posterior when the event is localized in an area where the galaxy catalog has a low completeness fraction which happens for most of the GW events that we consider in this analysis.
On the other hand, population assumptions are less important for events with a small localization in a region of the galaxy catalog that is complete. In these cases (for example GW190814), the posterior is dominated by the in-catalog likelihood terms and hence exhibit a weak dependence on the population assumptions.

The aforementioned discussion also explains the difference between our results and those found in \citet{Finke:2021aom} using GWTC--2 events. In that work the authors found a weak dependence of their posterior on the source population parameters. However, in their case only a few events, above a given completeness threshold of 70\% for the main result, were used to explore systematic effects due to the source population. Moreover, in exploring these systematics they varied the population assumptions only within the range of uncertainties reported in \citet{Abbott:2020gyp}, which already assumes a fixed cosmological model with a value of $H_0$ consistent with the Planck results \citep{Planck:2015fie}. Consequently, the results obtained by them are primarily driven by the Planck cosmological parameters. 

Finally, we also explore the systematics introduced by choices related to the galaxy catalog data.  In Fig.~\ref{fig:gwcosmo_result_sys} we also show the $H_0$ posteriors obtained with the \gladeplus catalog, but using K--band galaxies without luminosity weighting and $B_J$--band galaxies with luminosity weighting. In both cases, the $H_0$ posterior is not significantly affected by this choice and it is, again, dominated by the population assumptions. 

However, the impact of using luminosity weights is not negligible. For instance, in the case of GW190814 (see Fig.~\ref{fig:posteriors_pop}), removing the extra luminosity weight suppresses the $H_0$ posterior peak around $70 \hu$. This arises because the luminous galaxies shown in Fig.~\ref{fig:GW190814A_overdensity}, observed in the K--band, are now contributing to the GW event redshift localization with the same probability as the other 200+ galaxies included in the GW localization volume. 
We have verified that the $H_0$ posterior with the K--band and using luminosity weights does not depend on the faint end magnitude limit used for the analysis.

\subsection{Additional systematic uncertainties}
As we have seen, the presence of a known source mass scale can be used to measure cosmological parameters \citep{Farr:2019twy, 2021PhRvD.104f2009M}. However, if the source mass distribution is mismodeled, then the cosmological inference will be biased. With the current set of events, this effect contributes the dominant source of systematic uncertainty in the measurement of $H(z)$. In the population-based method, a key assumption is that the source mass distribution does not evolve with redshift~\citep{2021ApJ...912...98F}; any evolution is degenerate with the cosmological inference. Many BBH formation scenarios predict mild evolution in the mass distribution~\citep{2019MNRAS.487....2M,2021ApJ...907L..25W,2021arXiv210705702G,2021arXiv210906222M,2021arXiv211001634V}, but given our broad statistical uncertainties, we expect this evolution to only weakly affect our results, and we do not attempt to calibrate the mass models to theory. 
In the galaxy-catalog based method, we fix the source mass distribution. In this case, the choice of the peak location $\mu_{\rm g}$, associated with excess BHs in the distribution of source masses, is a main source of systematic uncertainty. If the $\mu_{\rm g}$ parameter is assumed to be lower (or higher) than its true value, this can lead to a higher (or lower) inferred value of $H_0$. The impact of this bias can be reduced if the support from the in-catalog part of the \textit{statistical galaxy catalog method} is informative, which is mainly possible for sources with better three-dimensional localization error and with a more complete galaxy catalog. In the future, as more GW detectors join the network, resulting in more events with better sky localizations, and galaxy catalogs which are more complete at high redshift, the impact of the source population on galaxy catalog method can be mitigated.

One of the additional sources of contamination, when seeking to infer the true luminosity distance $D_\mathrm{L}$ (and hence the true source masses) of a GW source in the absence of an EM counterpart, is the possible lensing of the GW signal due to the intervening matter distribution \citep{1992grle.book.....S, Bartelmann:2010fz, PhysRevLett.80.1138, Wang:1996as,Takahashi:2003ix, Dai:2016igl, Broadhurst:2018saj, Diego:2018fzr}. 
In the geometric optics limit, lensing modifies the GW signal by a magnification factor $\mu$ that only changes the amplitude of the GW strain, leading to a measured luminosity distance given by $\tilde{D}_\mathrm{L}= D_\mathrm{L}/\mu$. In the strong lensing limit, when the value of $\mu$ is large, the inferred luminosity distance to the source $\tilde{d}_L$ may be substantially lower than the true luminosity distance, i.e.\ introducing a bias in the measurement of the luminosity distance. However, for the GW detections considered here, the probability of observing such a strongly-lensed event is less than one percent \citep{Ng:2017yiu,Oguri:2018muv, Mukherjee:2020tvr}, even for a broad range of astrophysical time-delay scenarios \citep{Mukherjee:2021qam}. Moreover, searches from O1+O2 \citep{2019ApJ...874L...2H}, and more recent LIGO-Virgo analysis of the O3a data did not reveal any signs of strong lensing \citep{Abbott:2021iab}. Furthermore, searches for a stochastic GW background also provide a model-independent bound on the lensing event rate of BBHs \citep{Mukherjee:2020tvr,Buscicchio:2020cij,Abbott:2021iab}, which is again consistent with a low probability of contamination due to strong lensing in the GW sources considered here. In this paper we, therefore, ignore any possible impact of strong lensing on our cosmological parameter estimates. 

Apart from strong lensing, weak lensing of GW sources can also be a potential source of contamination. However, due to the effects of sky averaging, weak lensing should not produce a bias in the inferred values of the cosmological parameters, but will introduce additional variance on the luminosity distance of individual sirens at the level of a few percent \citep{Holz:1997ic, 2010PhRvD..81l4046H} and it is sub-dominant in comparison to the intrinsic measurement error (of about $20\%$) on the luminosity distance even for the best source in our current GW catalog. Hence, we also ignore the uncertainty due to weak lensing in this paper. However, in a future analysis, we could include the contribution from weak lensing and its impact on the distance measurement \citep{Holz:1997ic, Cutler:2009qv,  2010PhRvD..81l4046H, Namikawa:2015prh, Mukherjee:2019wcg}.

\section{Conclusions \label{sec:conclusions}}

Using the \Nevgwcosmo  GW events with detected SNR $>11$ reported in the third LIGO-Virgo-KAGRA Gravitational-Wave Transient Catalog \citep{O3bCat_PLACEHOLDER}, we have inferred constraints on the cosmological parameters adopting two different approaches: hierarchical inference without galaxy surveys and the statistical galaxy catalog method. We present for the first time analysis that constrains jointly the properties of the population of BBHs and the parameters of the cosmological model, and we have shown the crucial correlation that exists between the two sectors.

We have shown that an excess population of BHs in the mass range $30$ -- $40\, \Msol$, pointed out by \citet{Abbott:2020gyp}, is robust to the choice of assumed cosmological model parameters. While our constraints on the present-day matter density, $\Omega_{\rm m}$, and dark energy EoS, $w_0$, parameters are weak, we have measured the Hubble constant to be $H_0=\HnotnoticarogwwithBNSlogPLG\, \hu$ at $68\%$ CL from combining dark sirens with information from the bright siren GW170817 and its electromagnetic counterpart \citep{Abbott:2018wiz}. This result represents an improvement of \HnoticarogwimproGWTConePLG\% with respect to the $H_0$ value reported from analysis of O1 and O2 data \citep{Abbott:2019yzh} that made use of galaxy catalogs alone to infer statistical redshift information. In our analysis we also obtain weak constraints on the expansion history as a function of redshift.  

In addition we provide a constraint on the value of $H_0$ adopting a fixed \PLG\, population model of BBHs and using statistical redshift information inferred from the \gladeplus galaxy catalog. This analysis obtained, for the K--band, $H_0 = \HnotgwcosmoKband\, \hu$, which represents an improvement of \HnotgwcosmoimproGWTConeKband \% with respect to \citet{Abbott:2019yzh} alone, and an improvement of 20\% with respect to recent $H_0$ studies using GWTC--2 events \citep{Finke:2021aom}.  Most of the constraining power in our $H_0$ inference comes from the event GW170817 using its electromagnetic counterpart.  Combining the above result with information from GW170817 we obtain $\HnotgwcosmoBNSKband \hu$.
The most informative dark siren in the GWTC--3 catalog is GW190814, which alone provides an estimate of $H_0=\HnotgwcosmoGWnof\, \hu$, provided that the luminosity weighting scheme is applied.

A summary of the different $H_0$ values obtained  using different data sets and model assumptions can be seen in Table ~\ref{tab:recap}. The table is divided into two parts. The first part summarises the values that we infer for $H_0$ when fixing the population model to the most favorable one and then varying the luminosity band from \gladeplus\ used in our analysis. We used both the $B_J$ band and the K band and the results are very similar (see also Fig.~\ref{fig:gwcosmo_result_sys}). The second part of the table summarises the results obtained by marginalizing over the population parameters and using no galaxy catalog information. 

\begin{table*}[t]
  \centering
  \begin{tabular}{p{4.5cm}>{\centering\arraybackslash}p{2.5cm}>{\centering\arraybackslash}p{3.8cm}>{\centering\arraybackslash}p{2.5cm}>{\centering\arraybackslash}p{2.5cm}}
    \hline
    \centering {\bf Description} &\centering \textbf{Galaxy catalog} &\centering \textbf{BBH mass model} & \centering \boldmath$H^{\rm HDI}_{0}$ &  \boldmath$H^{\rm sym}_{0}$ \\
     & & & $\mathbf{[ km \ s^{-1} \ Mpc^{-1}]}$ & $\mathbf{[ km \ s^{-1} \ Mpc^{-1}]}$ \\ 
    \hline
    \hline
    \multirow{3}{*}{\parbox{4.5cm}{No galaxy catalog, Marginalizing over population model, $\Nevicarogw$ events}} & \centering - & \centering \PL &  $\HnoticarogwPL\,\,\,
    (\HnotnoticarogwwithBNSlogPL)$ &   $\HnoticarogwSYMPL\,\,\,
    (\HnotnoticarogwwithBNSlogSYMPL)$  \\
    & \centering - & \centering \PLG & $\HnoticarogwPLG\,\,\,(\HnotnoticarogwwithBNSlogPLG)$& $\HnoticarogwSYMPLG\,\,\,(\HnotnoticarogwwithBNSlogSYMPLG)$  \\
    & \centering - & \centering \BPL &  $\HnoticarogwBPL\,\,\,(\HnotnoticarogwwithBNSlogBPL)$ & $\HnoticarogwSYMBPL\,\,\,(\HnotnoticarogwwithBNSlogSYMBPL)$  \\
    \hline
    \hline
    \multirow{2}{*}{\parbox{4.5cm}{Using galaxy catalog, Fixed population model, $\Nevgwcosmo$ events}}&\centering \gladeplus K--band  & \centering \PLG & $\HnotgwcosmoKband \,\,\, (\HnotgwcosmoBNSKband)$ & $\HnotgwcosmoSYMKband\,\,\, (\HnotgwcosmoBNSSYMKband)$  \\
    &\centering \gladeplus $B_J$--band  & \centering \PLG &  $\HnotgwcosmobJband \,\,\, (\HnotgwcosmoBNSbJband)$ &  $\HnotgwcosmoSYMbJband \,\,\, (\HnotgwcosmoBNSSYMbJband)$  \\
    \hline
    \hline
  \end{tabular}
  \caption{Values of the Hubble constant obtained in this study using different data sets and analysis methods. The columns are in order: short description of the sources used in the study with SNR$>11$; galaxy catalog used (where appropriate); BBH mass model used and the $68.3\%$ CL $H_0$ value. The last two columns report the median and symmetric 90\% CI $H_0$ values. The values in the parenthesis is that obtained after combining with the GW170817 EM counterpart posterior.}
  \label{tab:recap}
\end{table*}

Although we have improved our previously reported constraints on the value of $H_0$ using these $\Nevgwcosmo$ GW events, our results are still dominated by the systematic effects induced by the assumptions made about the GW source population. The choice of mass scale set by $\mu_{\rm g}$, the mass at which the excess of BHs is centered, plays a crucial role in constraining the value of $H_0$. 


In the future, with significantly more both bright and dark sirens it will be possible to make robust measurements of $H_0$ and other cosmological parameters. On the one hand, measurement of bright sirens will help greatly with inferring the redshift from direct observations of EM counterparts \citep{Holz:2005df,Dalal:2006qt,  2010ApJ...725..496N, Chen:2017rfc,Feeney:2018mkj}. On the other hand, for dark sirens, the application of cross-correlation techniques to infer the clustering redshift of GW sources \citep{2021PhRvD.103d3520M, Bera:2020jhx} using spectroscopic galaxy surveys \citep{Diaz:2021pem}, the PISN mass scale of black holes \citep{Farr:2019twy, Mastrogiovanni:2020mvm}, the redshift distribution of the GW  sources \citep{Ding:2018zrk, Ye:2021klk}, and the tidal distortion of neutron stars \citep{2012PhRvL.108i1101M, Chatterjee:2021xrm} will enable robust measurement of the cosmic expansion history. With the aid of more observations and further development of analysis techniques, we will be able to reduce the current systematics and proceed towards accurate and precision gravitational-wave cosmology.  
\section*{Acknowledgments}
This material is based upon work supported by NSF’s LIGO Laboratory which is a major facility
fully funded by the National Science Foundation.
The authors also gratefully acknowledge the support of
the Science and Technology Facilities Council (STFC) of the
United Kingdom, the Max-Planck-Society (MPS), and the State of
Niedersachsen/Germany for support of the construction of Advanced LIGO 
and construction and operation of the GEO600 detector. 
Additional support for Advanced LIGO was provided by the Australian Research Council.
The authors gratefully acknowledge the Italian Istituto Nazionale di Fisica Nucleare (INFN),  
the French Centre National de la Recherche Scientifique (CNRS) and
the Netherlands Organization for Scientific Research (NWO), 
for the construction and operation of the Virgo detector
and the creation and support  of the EGO consortium. 
The authors also gratefully acknowledge research support from these agencies as well as by 
the Council of Scientific and Industrial Research of India, 
the Department of Science and Technology, India,
the Science \& Engineering Research Board (SERB), India,
the Ministry of Human Resource Development, India,
the Spanish Agencia Estatal de Investigaci\'on (AEI),
the Spanish Ministerio de Ciencia e Innovaci\'on and Ministerio de Universidades,
the Conselleria de Fons Europeus, Universitat i Cultura and the Direcci\'o General de Pol\'{\i}tica Universitaria i Recerca del Govern de les Illes Balears,
the Conselleria d'Innovaci\'o, Universitats, Ci\`encia i Societat Digital de la Generalitat Valenciana and
the CERCA Programme Generalitat de Catalunya, Spain,
the National Science Centre of Poland and the European Union – European Regional Development Fund; Foundation for Polish Science (FNP),
the Swiss National Science Foundation (SNSF),
the Russian Foundation for Basic Research, 
the Russian Science Foundation,
the European Commission,
the European Social Funds (ESF),
the European Regional Development Funds (ERDF),
the Royal Society, 
the Scottish Funding Council, 
the Scottish Universities Physics Alliance, 
the Hungarian Scientific Research Fund (OTKA),
the French Lyon Institute of Origins (LIO),
the Belgian Fonds de la Recherche Scientifique (FRS-FNRS), 
Actions de Recherche Concertées (ARC) and
Fonds Wetenschappelijk Onderzoek – Vlaanderen (FWO), Belgium,
the Paris \^{I}le-de-France Region, 
the National Research, Development and Innovation Office Hungary (NKFIH), 
the National Research Foundation of Korea,
the Natural Science and Engineering Research Council Canada,
Canadian Foundation for Innovation (CFI),
the Brazilian Ministry of Science, Technology, and Innovations,
the International Center for Theoretical Physics South American Institute for Fundamental Research (ICTP-SAIFR), 
the Research Grants Council of Hong Kong,
the National Natural Science Foundation of China (NSFC),
the Leverhulme Trust, 
the Research Corporation, 
the Ministry of Science and Technology (MOST), Taiwan,
the United States Department of Energy,
and
the Kavli Foundation.
The authors gratefully acknowledge the support of the NSF, STFC, INFN and CNRS for provision of computational resources.

This work was supported by MEXT, JSPS Leading-edge Research Infrastructure Program, JSPS Grant-in-Aid for Specially Promoted Research 26000005, JSPS Grant-in-Aid for Scientific Research on Innovative Areas 2905: JP17H06358, JP17H06361 and JP17H06364, JSPS Core-to-Core Program A. Advanced Research Networks, JSPS Grant-in-Aid for Scientific Research (S) 17H06133 and 20H05639 , JSPS Grant-in-Aid for Transformative Research Areas (A) 20A203: JP20H05854, the joint research program of the Institute for Cosmic Ray Research, University of Tokyo, National Research Foundation (NRF) and Computing Infrastructure Project of KISTI-GSDC in Korea, Academia Sinica (AS), AS Grid Center (ASGC) and the Ministry of Science and Technology (MoST) in Taiwan under grants including AS-CDA-105-M06, Advanced Technology Center (ATC) of NAOJ, Mechanical Engineering Center of KEK.

{\it We would like to thank all of the essential workers who put their health at risk during the COVID-19 pandemic, without whom we would not have been able to complete this work.}

\bibliographystyle{aasjournal}
\bibliography{references}

\label{lastpage}

\appendix

\section{Population prior models \label{app:priors}}

\subsection{Models for background cosmologies}

We use a flat $\Lambda$CDM cosmological model with dark energy density as a function of redshift $z$ described by \citet{Linder:2002et}

\begin{equation}
    \rho_{\Lambda}(z)=\rho_{\Lambda,0}(1+z)^{3(1+w_0)},
\end{equation}
where $\rho_{\Lambda}$ is the dark energy density and $w_0$ is a phenomenological parameter. If the dark energy density is constant during the cosmic expansion, as it is in the standard cosmological model, then $w_0=-1$. The luminosity distance is calculated as

\begin{equation}
    D_\mathrm{L} (z) = \frac{c (1+z)}{H_0} \int_0^z \frac{{\rm d}z'}{\sqrt{\Omega_{\rm m}(1+z')^3+\Omega_{\Lambda,0}(1+z')^{3(1+w_0)}}},
\end{equation}
where $\Omega_{\rm m}$ and $\Omega_{\Lambda,0}$ are the present-day dimensionless matter and dark energy densities respectively and $\Omega_{\Lambda,0}=1-\Omega_{\rm m}$.

We consider two sets of priors for the cosmological background model which are indicated in Table ~\ref{tab:prior_c}.
For the first set of priors, we adopt a flat $\Lambda$CDM cosmology that restricts the Hubble constant to only the range compatible with the $H_0$ tension, while for the second set we adopt more general, wide priors.

\begin{table}[h!]
    \centering
    \begin{tabular}{ c p{11cm} p{2mm} p{3cm} }
        {} & \center{Restricted priors ($H_0$-tension)} & \\
        \hline
        {\bf Parameter} & \textbf{Description} &  & \textbf{Prior} \\\hline\hline
        $H_0$ & Hubble constant expressed in $\rm{km \, s^{-1}} \, Mpc^{-1}$ in the $H_0$-tension region. &  & $\mathcal{U}$($65$, $77$) \\
        $\Omega_{\rm m}$ & Present-day matter density of the Universe fixed to the mean value inferred from measurements of the CMB in \citet{Planck:2015fie} &  &  0.3065  \\
        $w_0$ & Dark energy equation of state parameter fixed to the value that corresponds to a constant density. &  & -1\\
        \hline
        \\
        {} & \center{Wide priors} & \\
        \hline
        {\bf Parameter} & \textbf{Description} &  & \textbf{Prior} \\\hline\hline
        $H_0$ & Hubble constant expressed in $\rm{km \, s^{-1}}\, Mpc^{-1}$ &  & $\mathcal{U}$($\Hnotmin$, $\Hnotmax$) \\
        $\Omega_{\rm m}$ & Present-day matter density of the Universe.&  &  $\mathcal{U}$($\Omnotmin$,$\Omnotmax$)  \\
        $w_0$ & Dark energy equation of state parameter. &  & $\mathcal{U}$($\wnotmin$,$\wnotmax$) \\
        \hline 
        \hline 
    \end{tabular}
    \caption{
    Summary of the priors on the cosmological parameters, for the two sets of priors considered.}
  \label{tab:prior_c}
\end{table}

\subsection{Merger rate and redshift distribution priors}

We model the binary merger rate using a phenomenological model introduced with the form of \citet{2014ARA&A..52..415M}, motivated by the fact that the binary formation rate might follow the star formation rate.
The parameterization that we use \citet{Callister:2020arv} for the merger rate in the detector frame is

\begin{equation}
    \frac{{\rm d} N}{{\rm d}t_{\rm d} {\rm d}z} = R_0 [1+(1+z_{\rm p})^{-\gamma - k}] \frac{\partial V_{\rm c}}{\partial z}(\Lambda_{\rm c}) \frac{1}{1+z}\, \frac{(1+z)^{\gamma}}{1+[(1+z)/(1+z_{\rm p)}]^{\gamma + k}} ,
    \label{eq:dN}
\end{equation}
where $R_0$ is the binary merger rate today, $V_{\rm c}$ is the comoving volume, $\gamma$ and $k$ are the slopes of the two power-law regimes before and after a turning point $z_{\rm p}$ and $\Lambda_{\rm c}$ are a set of parameters describing the cosmological expansion. The extra $1/(1+z)$ factor encodes the clock difference in the source and detector, while the factor $(1+z_{\rm p})$ ensures that today the merger rate is $R(z=0)=R_0$. The redshift prior, normalized over the redshift, can be expressed as

\begin{equation}
    \pi(z| \gamma, \kappa, z_{\rm p}, \Lambda_{\rm c}) = \frac{1}{C} \frac{{\rm d}N}{{\rm d}t_{\rm d} {\rm d}z},
\end{equation}
where  $C$ is the normalization factor calculated from Eq.~(\ref{eq:dN}). 

The priors that we use on the merger rate hyper-parameters are indicated in Table~\ref{tab:prior_r}. The prior ranges that we model are wide enough to include the effect of a possible time delay between the formation and the merger of the binary.

\begin{table}[h!]
    \centering
    \begin{tabular}{ c p{11cm} p{2mm} p{3cm} }
        \hline
        {\bf Parameter} & \textbf{Description} &  & \textbf{Prior} \\\hline\hline
        $R_0$& BBH merger rate today in Gpc$^{-3}$ yr$^{-1}$&  & $\mathcal{U}$($\Rnotmin$, $\Rnotmax$) \\
        $\gamma$ & Slope of the powerlaw regime for the rate evolution before the point $z_p$ &  & $\mathcal{U}$($\gammamin$, $\gammamax$) \\
        $k$ & Slope of the powerlaw regime for the rate evolution after the point $z_{\rm p}$ &  & $\mathcal{U}$($\kappamin$, $\kappamax$)\\
        $z_p$ & Redshift turning point between the powerlaw regimes with $\gamma$ and $k$ &  & $\mathcal{U}$($\zpmin$, $\zpmax$) \\
        \hline
        \hline 
    \end{tabular}
    \caption{
    Summary of the prior hyper-parameters used for the merger rate evolution models adopted in this paper.}
  \label{tab:prior_r}
\end{table}

\subsection{Phenomenological Mass priors}

The three phenomenological mass models that we implement are a superposition of two probability density distributions and are compatible with the phenomenological priors used for BBHs in \citet{Abbott:2020gyp, O3bPops_PLACEHOLDER}, although the prior ranges on the population parameters are different. The first is a truncated power law $\mathcal{P}(x|x_{\rm min},x_{\rm max},\alpha)$ described by slope $\alpha$, and lower and upper bounds $x_{\rm min},x_{\rm max}$ at which there is a hard cut-off

\begin{equation}
\mathcal{P}(x|x_{\rm min},x_{\rm max},\alpha) \propto 
\begin{cases}
    x^\alpha & \left(x_{\rm min}\leqslant x \leqslant x_{\rm max}\right) \\
    0 & \mathrm{Otherwise}.
\end{cases}
\end{equation}
The second is a Gaussian distribution with  mean $\mu$ and standard deviation $\sigma$,

\begin{equation}
\mathcal{G}(x|\mu,\sigma,a,b)=\frac{1}{\sigma\sqrt{2\pi}} \exp{\left[ -\frac{(x-\mu)^2}{2\sigma^2}
\right]}.
\end{equation}
The source mass priors for the BBHs population that we consider are factorized as 

\begin{equation}
\pi(m_{1},m_{2}|\Phi_m)=\pi(m_{1}|\Phi_m)\pi(m_{2}|m_{1},\Phi_m),
\end{equation}
where $\pi(m_{1}|\Phi_m)$ is the distribution of the primary mass component while $\pi(m_{2}|m_{1},\Phi_m)$ is the distribution of the secondary mass component given the primary.
For all of the mass models, the secondary mass component $m_{2}$ is described with a truncated power-law with slope $\beta$ between a minimum mass $\mmin$ and a maximum mass $m_{1}$

\begin{equation}
    \pi(m_{2}|m_{1},\mmin,\alpha)=\mathcal{P}(m_{2}|\mmin,m_{1},\beta),
\end{equation}
while the primary mass is described with several models discussed in the following paragraphs.

For some of the phenomenological models, we also apply a smoothing factor to the lower end of the mass distribution
\begin{equation}
\pi(m_{1},m_{2}|\Phi_m)=[\pi(m_{1}|\Phi_m)\pi(m_{2}|m_{1},\Phi_m)]S(m_1|\delta_m,\mmin)S(m_2|\delta_m,\mmin),
\end{equation}
where $S$ is a sigmoid-like window function that adds a tapering of the lower end of the mass distribution. See Eq.~(B6) and Eq.~(B7) of \citet{Abbott:2020gyp} for the explicit expression for the window function. 

The three phenomenological mass models are highlighted in the following list. In Table~\ref{tab:priors}, we report the prior ranges used for the population hyper-parameters.

\begin{itemize}
    \item \textbf{\PL model}:  It describes the distribution of the primary mass $m_{1}$ with a truncated power law with slope $-\alpha$ between a minimum mass $\mmin$ and a maximum mass $\mmax$. 
\begin{equation}
    \pi(m_{1}|\mmin,\mmax,\alpha)=\mathcal{P}(m_{1}|\mmin,\mmax,-\alpha) .
\end{equation}

    \item \textbf{\PLG model}: It describes the primary mass component as a superposition of a truncated PL, with slope $-\alpha$ between a minimum mass $\mmin$ and a maximum mass $\mmax$, plus a Gaussian component with mean $\mu_{\rm g}$ and standard deviation
    $\sigma_{\rm g}$,
\begin{equation}
    \pi(m_{1}|\mmin,\mmax,\alpha,\lambda_{\rm g},\mu_{\rm g},\sigma_{\rm g})=[(1-\lambda_{\rm g})\mathcal{P}(m_{1}|m_{\rm min},m_{\rm max},-\alpha)+\lambda_{\rm g} \mathcal{G}(m_{1}|\mu_{\rm g},\sigma_{\rm g})].
    \label{eq:PLG}
\end{equation}
    
    \item  \textbf{\BPL model}: It describes the distribution of $m_{1}$ as a PL between a minimum mass $\mmin$ and a maximum mass $\mmax$. The \BPL model is characterized by two PL slopes $\alpha_1$ and $\alpha_2$ and by a breaking point between the two regimes at $m_{\rm break}=b (\mmax-\mmin)$, where $b$ is a number $\in [0,1]$. The broken PL model is 

\begin{equation}
    \pi(m_{1}|\mmin,\mmax,\alpha_1,\alpha_2)=\mathcal{P}(m_{1}|\mmin,m_{\rm break},-\alpha_1)+\frac{\mathcal{P}(m_{\rm break}|\mmin,m_{\rm break},-\alpha_1)}{\mathcal{P}(m_{\rm break}|m_{\rm break},\mmax,-\alpha_2)}\mathcal{P}(m_{1}|b,m_{\rm max},-\alpha_2).
\end{equation}

\end{itemize}

\begin{table}[h!]
    \centering
    \begin{tabular}{ c p{10cm} p{2mm} p{4cm} }
        {} & \centering{\PL} & \\
        \hline
        {\bf Parameter} & \textbf{Description} &  & \textbf{Prior} \\\hline\hline
        $\alpha$ & Spectral index for the PL of the primary mass distribution. &  & $\mathcal{U}$($\PLalphamin$, $\PLalphamax$) \\
        $\beta$ & Spectral index for the PL of the mass ratio distribution. &  & $\mathcal{U}$($\PLbetamin$, $\PLbetamax$) \\
        $\mmin$ & Minimum mass of the PL component of the primary mass distribution. &  & $\mathcal{U}$($\PLmminmin \, \Msol$, $\PLmminmax \, \Msol$)\\
        $\mmax$ &  Maximum mass of the PL component of the primary mass distribution. &  & $\mathcal{U}$($\PLmmaxmin\, \Msol$, $\PLmmaxmax\, \Msol$)\\
        \hline
        \\
        {} & \centering{\PLG} & \\
        \hline
        {\bf Parameter} & \textbf{Description} &  & \textbf{Prior} \\\hline\hline
        $\alpha$ & Spectral index for the PL of the primary mass distribution. &  & $\mathcal{U}$($\PLGalphamin$, $\PLGalphamax$)\\
        $\beta$ & Spectral index for the PL of the mass ratio distribution. &  & $\mathcal{U}$($\PLGbetamin$, $\PLGbetamax$)\\
        $\mmin$ & Minimum mass of the PL component of the primary mass distribution. &  & $\mathcal{U}$($\PLGmminmin\, \Msol$, $\PLGmminmax\, \Msol$)\\
        $\mmax$ &  Maximum mass of the PL component of the primary mass distribution. &  & $\mathcal{U}$($\PLGmmaxmin\, \Msol$, $\PLGmmaxmax\, \Msol$)\\
        $\lambda_{\rm g}$ & Fraction of the model in the Gaussian component. &  & $\mathcal{U}$($\PLGlambdapeakmin$, $\PLGlambdapeakmax$) \\
        $\mu_{\rm g}$ & Mean of the Gaussian component in the primary mass distribution.  &  & $\mathcal{U}$($\PLGmugmin\, \Msol$, $\PLGmugmax\, \Msol$) \\
        $\sigma_{\rm g}$ & Width of the Gaussian component in the primary mass distribution.  &  & $\mathcal{U}$($\PLGsigmagmin\, \Msol$, $\PLGsigmagmax\, \Msol$)\\
        $\delta_{m}$ & Range of mass tapering at the lower end of the mass distribution.  &  & $\mathcal{U}$($\PLGdeltammin\, \Msol$, $\PLGdeltammax\, \Msol$)\\
        \hline
        \\
        {} & \centering{\BPL} & \\
        \hline
        {\bf Parameter} & \textbf{Description} &  & \textbf{Prior} \\\hline\hline
        $\alpha_1$ & PL slope of the primary mass distribution for masses below $m_\mathrm{break}$. &  & $\mathcal{U}$($\BPLalphaonemin$, $\BPLalphaonemax$)\\
        $\alpha_2$ & PL slope for the primary mass distribution for masses above $m_\mathrm{break}$. &  & $\mathcal{U}$($\BPLalphatwomin$, $\BPLalphatwomax$) \\
        $\beta$ & Spectral index for the PL of the mass ratio distribution. &  & $\mathcal{U}$($\BPLbetamin$, $\BPLbetamax$)\\
        $\mmin$ & Minimum mass of the PL component of the primary mass distribution. &  & $\mathcal{U}$($\BPLmminmin\, \Msol$, $\BPLmmaxmin\, \Msol$) \\
        $m_\mathrm{max}$ & Maximum mass of the primary mass distribution. &  & $\mathcal{U}$($\BPLmmaxmin\, \Msol$, $\BPLmmaxmax\, \Msol$) \\
        $b$ & The fraction of the way between $m_\text{min}$ and $m_\text{max}$ at which the primary mass distribution breaks. & &  $\mathcal{U}$($\BPLbmin$,$\BPLbmax$) \\
        $\delta_{m}$ & Range of mass tapering on the lower end of the mass distribution.  &  & $\mathcal{U}$($\BPLdeltammin\, \Msol$, $\BPLdeltammax\, \Msol$)\\
        \hline 
        \hline 
    \end{tabular}
    \caption{
    Summary of the priors used for the population hyper-parameters for the three phenomenological mass models.}
  \label{tab:priors}
\end{table}

\section{Full results from the population analysis and effect of different SNR cuts}
\label{app:fullicarogw}

We provide extra details on the joint inference of cosmological and population parameters using BBHs.

\begin{figure}[t]
    \hspace{-1.5cm}
    \includegraphics[scale=1.05]{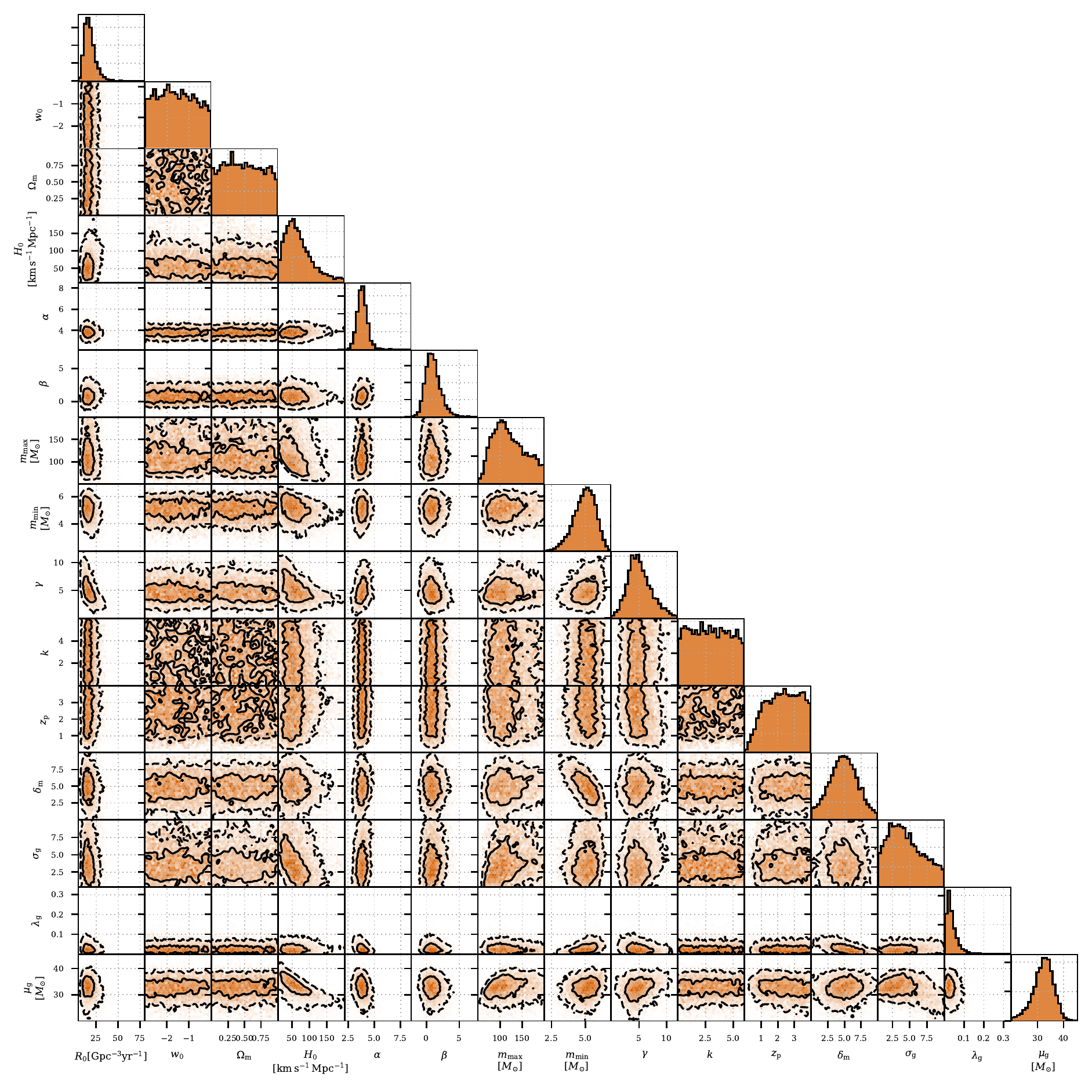}
    \caption{Corner plots for the preferred \PLG model parameters and cosmological parameters, fitted to BBHs with SNR $> 11$.}
    \label{fig:PLG_w0flatLCDM_corners_full}
\end{figure}

In Fig.~\ref{fig:PLG_w0flatLCDM_corners_full} we show the corner plots for the posterior associated with the \PLG model (the one implemented for the galaxy catalog analysis) adopting wide priors on the cosmological parameters and for the population of BBHs.
As was also shown in Fig.~\ref{fig:PLG_w0flatLCDM_corners}, the main mass-related population parameters correlating with the cosmological parameters, and in particular $H_0$, are the position of the Gaussian component (BBH excess in the mass distribution) and the higher end of the source mass distribution.

The other parameter that correlates with the estimation of $H_0$ is the rate evolution parameter $\gamma$. This parameter models a power-law increasing merger rate with the redshift. We find that higher values of $\gamma$ support lower values of $H_0$, which is due to the fact that lowering $H_0$ will place events at lower redshifts, which are incompatible with the observed mass distribution; therefore $\gamma$ tries to correct this by supporting higher redshifts. However, the posterior on the rate evolution is well within the statistical uncertainties given in \citep{O3bPops_PLACEHOLDER}.

We  run additional systematic studies using different SNR cuts for the selection of the BBHs to include in our analysis. We explore a higher SNR cut of 12 (more pure) that selects a sample of $\NevicarogwSNRtwelve$ events.
We also explore a lower SNR cut of 10, allowing $\NevicarogwSNRten$ events with a IFAR$>0.5$ yr and no plausible instrumental origin. For this set of events, GW190426\_152155 and  GW190531\_023648 are excluded as their secondary mass extends to lower masses in the NS region.

\begin{figure}
    \centering
    \includegraphics{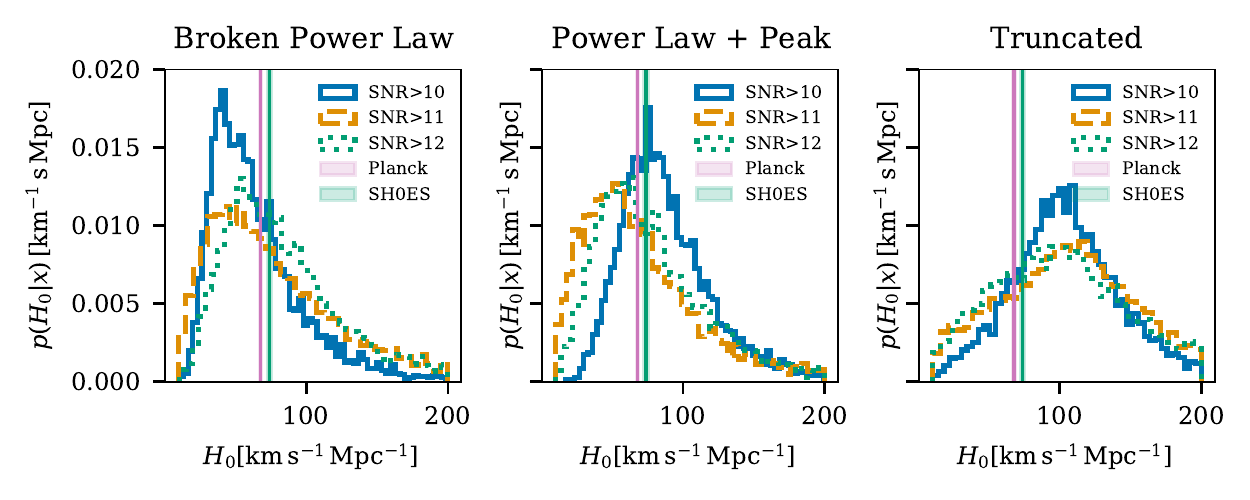}
    \caption{Marginal posterior probability distributions for $H_0$ with using BBH events with three different SNR cuts and the three mass models.}
    \label{fig:H0_comparison_SNR}
\end{figure}

The marginal $H_0$ posterior for all the mass models is shown in Fig.~\ref{fig:H0_comparison_SNR}, where we show that including more events always produces a posterior on the $H_0$ within the statistical uncertainties of other selection criteria. In all of these cases, the excess of BBHs around $35 \Msol$ is present for all the SNR cuts and it is responsible for the preference observed in the $H_0$ posterior.

\section{Schechter luminosity function studies}
\label{app:catalogsystematics}

In this Appendix we show comparisons of the Schechter luminosity function (LF) for the galaxies with the K and $B_J$ bands galaxies reported in \gladeplus.
Wrong assumptions on the LF, or incorrect description of the selection biases, could potentially introduce a bias in the inferred $H_0$.
Indeed, one of the key assumptions that we have made to construct our completeness corrections is that the galaxy catalog is \emph{magnitude limited}, i.e. galaxies are not detected only because they are too faint. This cannot be the case if another selection bias (based on e.g., colors or spectral features) were present, or if color and evolution corrections are not implemented properly.

\begin{figure}
    \centering
    \includegraphics{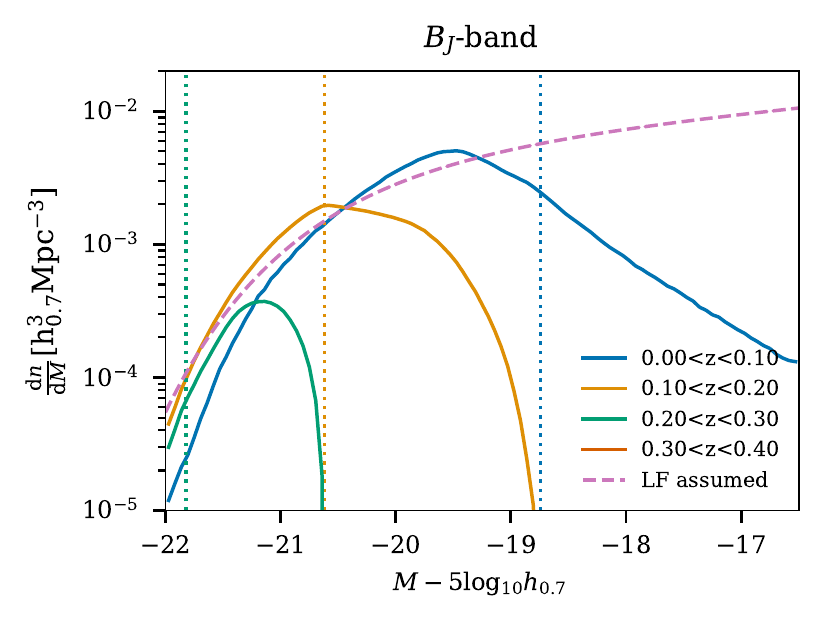}
    \includegraphics{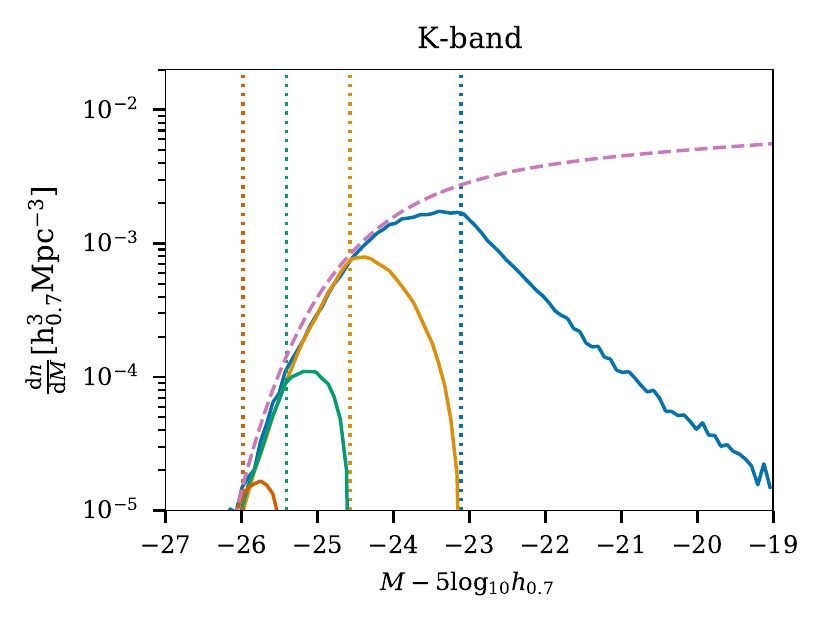}
    \caption{\textit{Left:} Comparison of the assumed $B_J$--band galaxy luminosity function (pink dashed line) and the differential number density of galaxies in different redshift bins (solid colored lines). The vertical dashed line indicates the median absolute magnitude threshold for galaxy detection, as computed in our code. \textit{Right:} Same comparison, but for the K--band luminosity function. In each panel the solid lines are calculated based on the median apparent magnitude threshold, computed over all sky directions, then  collecting and making a histogram of all galaxies in the \gladeplus\ catalog that lie in the appropriate redshift bin and are brighter than the apparent magnitude threshold.}
    \label{fig:LF_studies}
\end{figure}

In Fig.~\ref{fig:LF_studies} we show a comparison of the assumed LF and the number density of galaxies per comoving volume present in \gladeplus. In the case that the galaxy catalog can be correctly described as magnitude-limited, we expect that the distribution of the \gladeplus galaxies distribution will match the assumed LF at its bright end, and then will start to decrease when we reach (and exceed) the corresponding absolute magnitude threshold. Galaxies in the K--band are well described by this behaviour and missing galaxies can be explained by the impact of the apparent magnitude threshold, while for the $B_J$--band there seems to be some additional missing galaxies at low redshift. This observed behaviour motivated our decision to present our main results using the K--band magnitudes compiled in the \gladeplus catalog.

\end{document}